\documentclass[12pt]{article}
\usepackage{epsfig}
\usepackage{epstopdf}
\usepackage{hyperref}
\hypersetup{ colorlinks=true, linktoc=all, linkcolor=blue}
\usepackage{makeidx}
\usepackage{amssymb}
\usepackage{amsmath}
\usepackage{algorithm}
\usepackage{algorithmic}
\usepackage{datetime}
\usepackage{tikz-feynman}
\usepackage{adjustbox}
\makeindex
\usepackage[top=2cm,bottom=2cm]{geometry} 
\begin{document}

\newcommand{\nc}{\newcommand}
\nc{\noi}{\noindent}
\newtheorem{opgave}{Excercise}
\nc{\bop}[1]{\begin{opgave}\begin{rm}{\bf#1} \newline\noi}
\nc{\eop}{\end{rm}\end{opgave}}
\nc{\dia}[1]{\mbox{{\Large{\bf diag}}}[#1]}
\nc{\diag}[3]{\raisebox{#3cm}{\epsfig{figure=diagrams/d#1.eps,width=#2cm}}}
\nc{\dg}[4]{\raisebox{#3cm}{\epsfig{figure=diagrams/album/#1.eps,width=#2cm}}\;\;\mbox{\small$#4$}\hspace{1cm}}
\nc{\dgs}[3]{\raisebox{#3cm}{\epsfig{figure=diagrams/album/#1.eps,width=#2cm}}}
\nc{\dgw}[3]{\raisebox{#3cm}{\epsfig{figure=diagrams/album/#1.eps,width=#2cm}}}
\nc{\bq}{\begin{equation}}
\nc{\eq}{\end{equation}}
\nc{\bqa}{\begin{eqnarray}} 
\nc{\eqa}{\end{eqnarray}}
\nc{\eqan}{\nonumber\end{eqnarray}}
\nc{\nl}{\nonumber \\}
\nc{\grf}{Green's function}
\nc{\grfs}{Green's functions}
\nc{\cgrf}{connected Green's function}
\nc{\cgrfs}{connected Green's functions}
\nc{\f}{\varphi}
\nc{\exb}[1]{\exp\!\left(#1\right)}
\nc{\avg}[1]{\left\langle #1\right\rangle}
\nc{\suml}{\sum\limits}
\nc{\prodl}{\prod\limits}
\nc{\intl}{\int\limits}
\nc{\ddv}[1]{{\partial\over\partial #1}}
\nc{\ddvv}[2]{{\partial^{#1}\over\left(\partial #2\right)^{#1}}}
\nc{\la}{\lambda}
\nc{\La}{\Lambda}
\nc{\eqn}[1]{Eq.(\ref{#1})}
\nc{\eqns}[1]{Eqs.(\ref{#1})}
\nc{\appndix}[3]{\section{{#2}\label{#3}}\input{#3}}
\nc{\cala}{{\cal A}}
\nc{\calb}{{\cal B}}
\nc{\cale}{{\cal E}}
\nc{\calr}{{\cal R}}
\nc{\calf}{{\cal F}}
\nc{\calh}{{\cal H}}
\nc{\calc}{{\cal C}}
\nc{\cald}{{\cal D}}
\nc{\calp}{{\cal P}}
\nc{\calt}{{\cal T}}
\nc{\calk}{{\cal K}}
\nc{\cals}{{\cal S}}
\nc{\caln}{{\cal N}}
\nc{\cali}{{\cal I}}
\nc{\calq}{{\cal Q}}
\nc{\calg}{{\cal G}}
\nc{\calphi}{\Psi}
\nc{\lra}{\;\leftrightarrow\;}
\nc{\stapel}[1]{\begin{tabular}{c} #1\end{tabular}}
\nc{\kolom}[1]{\left(\stapel{#1}\right)}
\nc{\al}{\alpha}
\nc{\be}{\beta}
\nc{\g}{\gamma}
\nc{\ga}{\gamma}
\nc{\Ga}{\Gamma}
\nc{\ka}{\kappa}
\nc{\om}{\omega}
\nc{\si}{\sigma}
\nc{\Si}{\Sigma}
\nc{\Sibar}{\overline{\Si}}
\nc{\ro}{\rho}
\nc{\tha}{\theta}
\nc{\ze}{\zeta}
\nc{\vph}{\vphantom{{A^A\over A_A}}}
\nc{\wph}{\vphantom{{A^A}}}
\nc{\order}[1]{{\cal O}\left(#1\right)}
\nc{\De}{\Delta}
\nc{\de}{\delta}
\nc{\vx}{\vec{x}}
\nc{\vxi}{\vec{\xi}}
\nc{\vy}{\vec{y}}
\nc{\vk}{\vec{k}}
\nc{\vn}{\vec{n}}
\nc{\vq}{\vec{q}}
\nc{\vp}{\vec{p}}
\nc{\vz}{\vec{z}}
\nc{\vt}{\vec{t}}
\nc{\nv}{\vec{n}}
\nc{\ve}{\vec{e}}
\nc{\vs}{\vec{s}}
\nc{\vr}{\vec{r}}
\nc{\hk}{\hat{k}}
\nc{\lijst}[1]{\begin{center}
  \fbox{\begin{minipage}{12cm}{#1}\end{minipage}}\end{center}}
\nc{\lijstx}[2]{\begin{center}
  \fbox{\begin{minipage}{#1cm}{#2}\end{minipage}}\end{center}}
\nc{\eps}{\epsilon}
\nc{\caput}[2]{\chapter{#1}\input{#2}}
\nc{\pd}{\partial}
\nc{\dtil}{\tilde{d}}
\nc{\vskp}{\vspace*{\baselineskip}}
\nc{\coos}{co\-ord\-in\-at\-es}
\nc{\calm}{{\cal{M}}}
\nc{\calj}{{\cal J}}
\nc{\call}{{\cal L}}
\nc{\calu}{{\cal U}}
\nc{\calw}{{\cal W}}
\nc{\msq}{\langle|\calm|^2\rangle}
\nc{\nsym}{F_{\mbox{{\tiny{symm}}}}}
\nc{\dm}[1]{\mbox{\bf dim}\!\left[#1\right]}
\nc{\fourv}[4]{\left(\begin{tabular}{c}
  $#1$\\$#2$\\$#3$\\$#4$\end{tabular}\right)}
\nc{\driev}[3]{\left(\begin{tabular}{c}
  $#1$\\$#2$\\$#3$\end{tabular}\right)}
\nc{\fs}[1]{/\!\!\!#1}
\nc{\dbar}[1]{{\overline{#1}}}
\nc{\tr}[1]{\mbox{Tr}\left(#1\right)}
\nc{\deter}[1]{\mbox{det}\left(#1\right)}
\nc{\row}[4]{$#1$ & $#2$ & $#3$ & $#4$}
\nc{\matv}[4]{\left(\begin{tabular}{cccc}
   #1 \\ #2 \\ #3 \\ #4\end{tabular}\right)}
\nc{\twov}[2]{\left(\begin{tabular}{cccc}
   $#1$ \\ $#2$\end{tabular}\right)}
\nc{\ubar}{\dbar{u}}
\nc{\vbar}{\dbar{v}}
\nc{\wbar}{\dbar{w}}
\nc{\lng}{longitudinal}
\nc{\pol}{polarisation}
\nc{\longpol}{\lng\ \pol}
\nc{\bnum}{\begin{enumerate}}
\nc{\enum}{\end{enumerate}}
\nc{\nubar}{\overline{\nu}}
\nc{\mau}{{m_{\mbox{{\tiny U}}}}}
\nc{\mad}{{m_{\mbox{{\tiny D}}}}}
\nc{\qu}{{Q_{\mbox{{\tiny U}}}}}
\nc{\qd}{{Q_{\mbox{{\tiny D}}}}}
\nc{\vau}{{v_{\mbox{{\tiny U}}}}}
\nc{\aau}{{a_{\mbox{{\tiny U}}}}}
\nc{\aad}{{a_{\mbox{{\tiny D}}}}}
\nc{\vad}{{v_{\mbox{{\tiny D}}}}}
\nc{\mw}{{m_{\mbox{{\tiny W}}}}}
\nc{\mt}{{m_{\mbox{{\tiny t}}}}}
\nc{\mv}{{m_{\mbox{{\tiny V}}}}}
\nc{\mz}{{m_{\mbox{{\tiny Z}}}}}
\nc{\mf}{{m_{\mbox{{\tiny f}}}}}
\nc{\gz}{{\Gamma_{\mbox{{\tiny Z}}}}}
\nc{\mh}{{m_{\mbox{{\tiny H}}}}}
\nc{\gw}{{g_{\mbox{{\tiny W}}}}}
\nc{\gs}{{g_{\mbox{{\small s}}}}}
\nc{\qw}{{Q_{\mbox{{\tiny W}}}}}
\nc{\qe}{{Q_{\mbox{{\tiny e}}}}}
\nc{\qc}{{Q_{\mbox{{\tiny c}}}}}
\nc{\law}{{\Lambda_{\mbox{{\tiny W}}}}}
\nc{\obar}{\overline}
\nc{\gfer}{G_{\mbox{{\tiny F}}}}
\nc{\guuh}{{g_{\mbox{{\tiny UUH}}}}}
\nc{\gtth}{{g_{\mbox{{\tiny ttH}}}}}
\nc{\gffh}{{g_{\mbox{{\tiny ffH}}}}}
\nc{\gwwz}{{g_{\mbox{{\tiny WWZ}}}}}
\nc{\gwwh}{{g_{\mbox{{\tiny WWH}}}}}
\nc{\gwwhh}{{g_{\mbox{{\tiny WWHH}}}}}
\nc{\gzzh}{{g_{\mbox{{\tiny ZZH}}}}}
\nc{\gvvh}{{g_{\mbox{{\tiny VVH}}}}}
\nc{\ghhh}{{g_{\mbox{{\tiny HHH}}}}}
\nc{\ghhhh}{{g_{\mbox{{\tiny HHHH}}}}}
\nc{\gzzhh}{{g_{\mbox{{\tiny ZZHH}}}}}
\nc{\thw}{\theta_W}
\nc{\sw}{{s_{\mbox{{\tiny W}}}}}
\nc{\cw}{{c_{\mbox{{\tiny W}}}}}
\nc{\ward}[2]{\left.#1\right\rfloor_{#2}}
\nc{\vpa}{(1+\g^5)}
\nc{\vma}{(1-\g^5)}
\nc{\gwzh}{{g_{\mbox{{\tiny WZH}}}}}
\nc{\gudh}{{g_{\mbox{{\tiny UDH}}}}}
\nc{\gwcj}{{g_{\mbox{{\tiny Wcj}}}}}
\nc{\gwcjg}{{g_{\mbox{{\tiny Wcj$\g$}}}}}
\nc{\gwwcc}{{g_{\mbox{{\tiny WWcc}}}}}
\nc{\gzcc}{{g_{\mbox{{\tiny Zcc}}}}}
\nc{\none}[1]{ }
\nc{\Vir}{{V_{\mbox{{\tiny IR}}}}}
\nc{\kt}[1]{\left|#1\right\rangle}
\nc{\br}[1]{\left\langle#1\right|}
\nc{\bak}[3]{\left\langle #1\!\left| #2 \right|\! #3\right\rangle}
\nc{\bk}[2]{\left\langle #1\right|\left.\!\!#2\right\rangle}
\nc{\tw}{\tilde{w}}
\nc{\hDe}{\nabla}
\nc{\bino}[2]{\left(\begin{tabular}{c}$#1$\\$#2$\end{tabular}\right)}
\nc{\Om}{\Omega}
\nc{\Omb}{\overline{\Omega}}
\nc{\qbar}{\bar{q}}
\nc{\leeg}[1]{}
\nc{\dilog}{\mbox{Li}_2}
\nc{\marx}[1]{\marginpar{\fbox{\ref{#1}}}}
\nc{\ebar}{\bar{\eta}}
\nc{\eebar}{\eta\,\ebar}
\nc{\xibar}{\bar{\xi}}
\nc{\Sib}{\overline{\Si}}
\nc{\hw}{\hat{w}}
\nc{\cthe}{\cos\theta\;}
\nc{\sthe}{\sin\theta\;}
\nc{\ctthe}{\cos2\theta\;}
\nc{\stthe}{\sin2\theta\;}
\nc{\bmat}{\left(\begin{tabular}}
\nc{\emat}{\end{tabular}\right)}
\nc{\idx}[1]{\index{#1}#1}
\nc{\pars}{\partial_s}
\nc{\parv}{\partial_v}
\nc{\parw}{\partial_w}
\nc{\vv}{{\bf{v}}}
\nc{\vw}{{\bf{w}}}
\nc{\veps}{\vec{\eps}}
\nc{\vphi}{\vec{\phi}}
\nc{\fall}[2]{#1^{\underline #2}}
\nc{\sihad}{\si_{\mbox{{\tiny had}}}}
\nc{\dexy}{\de_{xy}}
\nc{\gotf}{\mathfrak{F}}
\nc{\gotm}{\mathfrak{m}}
\nc{\gf}{\mathfrak{f}}
\nc{\gh}{\mathfrak{h}}
\nc{\gc}{\mathfrak{c}}
\nc{\lagr}{{\mathfrak L}}
\nc{\gotc}{{\mathfrak C}}
\nc{\gotp}{{\mathfrak P}}
\nc{\got}{{\mathfrak T}}
\nc{\plaatje}[4]{\begin{figure}\begin{center}
  \includegraphics[width=#2cm]{#1}\caption{#3}\label{#4}\end{center}\end{figure}}
\nc{\figur}[1]{figure \ref{#1}}
\nc{\logbe}{\log\left({1+\be\over1-\be}\right)}
\nc{\calmlight}{\calm_{\mbox{\tiny{light}}}}
\nc{\calmlightcut}{\calm_{\mbox{\tiny{light,cut}}}}
\nc{\tinyf}{\begin{tiny}{\bf{\em F\/}}\end{tiny}}
\nc{\Veff}{V_{\mbox{{\tiny eff}}}}
\nc{\psibb}{\bar{\psi}}
\nc{\muh}{\hat{\mu}}
\nc{\lah}{\hat{\lambda}}
\nc{\uh}{\hat{u}}
\nc{\gow}{\mathfrak{W}}
\nc{\etp}{\eta_+}
\nc{\etm}{\eta_-}
\nc{\conto}{\mbox{{\Large $\leadsto$}}}
\nc{\sfc}[1]{\mbox{{\small ${#1}$}}}
\nc{\smf}{\mathfrak{sm}}
\nc{\vf}{\vec{\phi}}
\nc{\chib}{\bar{\chi}}
\nc{\vchi}{\vec{\chi}}
\nc{\vchib}{\vec{\chib}}
\nc{\hmu}{\hat{\mu}}
\nc{\hu}{\hat{u}}
\nc{\hal}{\hat{\al}}
\nc{\ran}{\mathfrak{r}()}
\nc{\vell}{\vec{\ell}}
\nc{\mtil}{\tilde{m}}
\nc{\msbar}{$\overline{\mbox{MS}}$}
\nc{\half}{\mbox{{\small${1\over2}$}}}
\nc{\drhalf}{\mbox{{\small${3\over2}$}}}
\nc{\kwart}{\mbox{{\small${1\over4}$}}}
\nc{\fc}{\phi_c}
\nc{\Pigg}{\Pi_{\g\g}}
\nc{\Piv}{\Pi_{\mbox{{\tiny V}}}}
\nc{\gt}{\mathfrak{t}}
\nc{\gn}{\mathfrak{n}}
\nc{\va}{\vec{a}}
\nc{\vb}{\vec{b}}
\nc{\vc}{\vec{c}}
\nc{\vd}{\vec{d}}
\nc{\vsi}{\vec{\si}}
\nc{\must}{\stackrel{\downarrow}{=}}
\nc{\vreps}{\varepsilon}
\nc{\cth}{c_\theta}
\nc{\sth}{s_\theta}
\nc{\da}{{\dot{a}}}
\nc{\db}{{\dot{b}}}
\nc{\dc}{{\dot{c}}}
\nc{\dd}{{\dot{d}}}
\nc{\dee}{{\dot{e}}}
\nc{\nosum}{\mbox{(no summation)}}
\nc{\Bbar}{{B}}
\nc{\one}{{\bf 1}}
\nc{\hA}{\hat{A}}
\nc{\hB}{\hat{B}}

\begin{center}
{\Large {\bf The Higgs Mechanism with Diagrams:\\a didactic approach}}\\
\vspace*{\baselineskip}
Jochem W. Kip and Ronald H.P. Kleiss\\ \vspace*{\baselineskip}
Institute for Mathematics, Astrophysics and Particle Physics\\
 Radboud University, Nijmegen, The Netherlands\\
\vspace*{\baselineskip}
{\it Dedicated to the memory of P.W. Higgs (1929-2024)}\\
\vspace*{3\baselineskip}
Abstract\\
\vspace*{\baselineskip}
\begin{minipage}{12cm}{
We present a p{\ae}dagogical treatment of the electroweak 
Higgs mechanism based solely on Feynman diagrams and S-matrix elements,
without recourse to (gauge) symmetry arguments.
Throughout, the emphasis is on Feynman rules and the Schwinger-Dyson equations;
it is pointed out that particular care is needed in the treatment of tadpole diagrams and their symmetry factors.
}
\end{minipage}
\end{center}
\newpage
\tableofcontents
\newpage
\section{Introduction}
\subsection{The diagrammatic approach}
In particle theory there exist two lines of thought that are well known, but are minority viewpoints. The first  is that
Feynman diagrams and their Feynman rules are a more fundamental  description of the physics than are Lagrangians
and actions \cite{diagrammar}. The second is that physical requirements like that of unitarity are more fundamental as 
restrictions on the form of a theory than symmetries that are imposed {\em a priori\/} \cite{CLT}. In this light,
it becomes interesting to see how the Higgs mechanism (based on spontaneous symmetry breaking (SSB),
although we will not explicitly use any symmetry arguments) can be cast into a diagrammatic
form without recourse to either Lagrangians or principles of gauge symmetry. This is what we shall explore.

What we shall ultimately derive is the electroweak standard model, so no {\em new\/} results are 
obtained. Rather, it is the {\em way\/} in which they are obtained that interests us here:
therefore we adopt a p{\ae}dagogical approach. We shall only consider either scattering amplitudes (S-matrix elements)
formed from Feynman diagrams with all external lines on-shell, or {\em off-shell\/} amplitudes in which {\em one\/}
line is kept off-shell.\footnote{Connecting two off-shell amplitude results in (a contribution to) an S-matrix element.}
The particular vertices of our
models are dictated by the requirement of unitarity, which in the case of massless vector particles
means current conservation. The `fields' of the theory are considered to be 
labels for bookkeeping of off-shell amplitudes, and particles take on their identity only upon LSZ truncation
of external lines \cite{LSZ}.
Feynman rules are proposed, not as following from an {\it a priori\/} gauge symmetry, but simply in order to 
make the theory unitary.\footnote{That is, we postulate Feynman rules instead of postulating a gauge theory. Inasmuch as
physics is the business of proposing rules, and then confront these with experiment, the two approaches are
methodologically equivalent.}

A possible stumbling block may be our use of {\em tachyons}. In a cosmologically inspired treatment of the Standard Model,
tachyons, that is modes with the `wrong' mass sign, crop up at a certain stage in the universe's 
cooling.
This is the point at which
SSB takes place, with the resulting particle spectrum being non-tachyonic. This is only possible because the would-be tachyons
are self-interacting in the famous `Mexican hat' potential: and that is what we will also use. The instability associated with
free tachyons is therefore irrelevant (see below).

Since we want to proceed didactically we shall move from simple to
more complicated models: therefore the layout of the paper is as follows. 
We start with a set of self-interacting {\em tachyons\/} (the Higgs sector)
that end up as a single massive scalar and a number of massless scalars. 
We then couple the tachyons to a massless
vector boson (the Abelian Higgs model) and see how the vector picks up a mass. 
Subsequently we extend the model to contain three self-interacting vectors (the Apollo model and the Higgs-Kibble model),
and then we add a single extra vector to arrive at the electroweak model. Finally, we discuss the inclusion of
fermions. A number of technical points are discussed in the appendices.\\

\subsection{On unitarity}
We have to specify what is meant by `unitarity' in this paper: we use that term for {\em partial-wave\/} unitarity. That is,
if in a given $n$-particle scattering amplitude we keep all angles fixed and let the overall energy scale $E$ grow much larger than
all masses,
the amplitude should asymptotically decrease as $E^{4-n}$ or faster. Loop corrections can modify this
behaviour by logarithmic terms at most, but such terms ought to come from exceptional values of the loop momenta, for instance
when these become very large, or very
collinear with other momenta. Should we restrict also the loop momenta to have magnitude of order $E$ and some
fixed direction,
then the asymptotic $E^{4-n}$ limit must be respected rigorously.

In a theory with scalar and vector propagators scaling as $E^{-2}$, with minimal coupling between the scalars and the
vectors, and with the usual self-interactions between the vectors,\footnote{Three-vector vertices going as $E^1$, and
four-vector vertices independent of $E$.} unitarity can be proven by simple
power counting as long as the external-line factors for the particles go as $E^0$. Problems arise if the
vectors are massive: then the appropriate, physical, `unitary-gauge' propagators\footnote{The unitary
gauge is sometimes, erroneously, described as non-renormalizable.} for the vectors scale asymptotically
as $E^0$, and an external vector's longitudinal polarization vector grows as $E^1$ since at large momentum $p$ 
it approaches $p/M$, where $M$ is the mass.
By showing that such a theory is identical to
one in which only massless vectors and tachyons occur, we thus prove that all our models,
including the electroweak model in the unitary gauge,
satisfy the unitarity requirement.

\section{Interacting tachyons}
\subsection{Higgs and Goldstones\label{getgold}}
We define a tachyon to have a {\em bare\/} scalar propagator with the `wrong' mass term. {\em Free\/} tachyons 
are  {\em physically\/} unacceptable
for reasons of causality, but as far as the Schwinger-Dyson equations (SDe) are concerned their {\em diagrammatic description\/}
is valid as long as the $i\eps$ term has the proper sign \cite{RK}. {\em Self-interacting\/} tachyons can be physically acceptable
as we shall see: the {\em eventual\/} particle spectrum is perfectly non-tachyonic.
We start with $N_t$ ($\ge2$) tachyons (labelled by $k$, $\ell$, $n,\ldots$) with the following Feynman rules:\footnote{We shall
 not explicitly write either $i\eps$ or $\hbar$ in what follows.}
\bqa
\diag{001}{1.2}{-.5} &=& {i\over q^2+m^2/2}\;\;,\nl
\diag{002}{1.2}{-.3} &=&
-i{m^2\over  v^2}T_{knpr}\;\;\;,\;\;\;T_{knpr} = \de_{kn}\de_{pr} + \de_{kp}\de_{rn} + \de_{kr}\de_{np}\;\;.
\eqa
The term $+m^2/2$ shows the tachyonic character, and $v$ parametrizes the strength of the self-interaction.
The tachyons may develop tree-level tadpoles $\tau_n$,\footnote{Since Lorentz invariance 
forbids tadpoles for non-zero spin, SSB requires the presence of scalar particles.} 
described by the SDe\footnote{In Lagrangian-speak, this is the Euler-Lagrange 
equation for vanishing momentum.}
\bq
\diag{003}{.8}{-.1} \;\;=\;\; \diag{004}{1.}{-.5}\;\;,\label{counter1}
\eq
which we can write out as
\bq
\tau_n = {2i\over m^2}\left({-3im^2\over v^2}{\tau_n^3\over3!}+\suml_{\ell\ne n}{-im^2\over v^2}{\tau_n\tau_\ell^2\over 2!}\right)
= {\tau_n\over v^2}\suml_\ell\tau_\ell^2\;\;,
\eq
so that either all $\tau_n$ vanish (the physically unacceptable solution), or $\tau_n = x_nv$ with $x_n$ the components of a unit vector in $N_t$-dimensional tachyon label space (tl-space):
\bq
\kt{\tau} = v\kt{x}\;\;\;,\;\;\;\bk{x}{x} = 1\;\;.
\eq
The tadpoles will dress the various propagators and mix them, since
\bq
\diag{005}{1.2}{-.1} \;= {-im^2\over v^2}\left(\vph\de_{nk}\left({3\tau_n^2\over2!}+\suml_{\ell\ne n}{\tau_\ell^2\over2!}\right)
+\left(1-\de_{kn}\right)\tau_n\tau_k\right) = -im^2\left(\half\de_{nk} + x_nx_k\right)\;\;.\label{passive1}
\eq
The dressed propagators have their own SDe:
\bq
\Pi = \diag{006}{1.4}{-.1} = \diag{007}{.8}{0} + \diag{008}{1.6}{-.1}\;\;.
\eq
Throughout this paper, hatched blobs stand for {\em connected\/} diagrams.
We give the following steps in detail, since we shall employ them again later on.
Multiplying by the denominator $(q^2+ \half m^2)$ we find
\bq
\Pi_{nk}(q^2+\half m^2) = i\de_{nk} + m^2\left(\wph\half\de_{n\ell} + x_nx_\ell\right)\Pi_{\ell k}\;\;.
\eq
Here and in the following, we employ the summation convention: {\em all\/} paired labels (in this case, $\ell$)
are to be summed over their appropriate range, unless specified otherwise.\footnote{It is under-appreciated that
the summation convention is naturally connected with the diagrammatic approach, since a line has precisely two
endpoints.} 
The terms with $\half m^2$ on either side cancel, so that
\bq
\Pi_{nk}\,q^2 = i\de_{nk} + m^2x_n\, x_\ell\Pi_{\ell k}\;\;.
\eq
Multiplying both sides by $x_n$ and summing over $n$, we find
\bq
 x_\ell\Pi_{\ell k} = {ix_k\over q^2-m^2}\;\;,\label{higgsresult}
\eq
and we arrive at
\bq
\Pi_{nk} = {i\over q^2-m^2}x_nx_k + {i\over q^2}\left(\de_{nk}-x_nx_k\right)\;\;.
\eq
We can choose a complete orthonormal basis in tl-space:
\bq
\bk{x}{x} = 1\;\;,\;\;\bk{y^j}{y^k} = \de_{jk}\;\;,\;\;\bk{x}{y^j} = 0\;\;,\;\;j,k\in\{2,\ldots,N_t\}\;\;,
\eq
so that 
\bq
\de_{nk}-x_nx_k = (y^j)_n(y^j)_k\;\;.
\eq
We now introduce the concept of {\em active\/} and {\em passive\/} vertices. The
active vertices are those in which at least two of the momenta
involved are linearly independent; the vertex of \eqn{passive1} is passive, not active. 
In every diagram that contributes to a scattering amplitude\footnote{This excludes vacuum diagrams; see below.}
any internal line must end in active vertices somewhere; 
therefore, concentrating on a particular internal line, we can write an amplitude as 
\bq
\calm = \diag{009}{2.4}{-.3} = A_n\Pi_{nk}B_k\;\;.
\eq
$A_n$ is the active off-shell amplitude emitting $n$, and $B_k$ is the one absorbing $k$.\footnote{The blobs $A$ and $B$
may be connected by other lines, in which case we have a loop diagram.} The active vertices are
indicated by dots. We see that we can write
\bq
\calm = A_h\,R_h(q)\,B_h + A_j\,R_0(q)\,B_j\;\;,
\eq
where 
\bq
R_h(q) ={i\over q^2-m^2}\;\;,\;\;R_0(q) = {i\over q^2}\;\;,\;\;A_h = A_nx_n\;\;,\;\;A_j = A_n(y^j)_n\;\;.
\eq
and similarly for $B$: the internal line therefore represents one massive propagator and a bunch of $N_t-1$ massless ones,
the so-called Goldstone bosons.
 By letting $A$ and $B$ move very far apart in spacetime, so that the internal line is truncated \cite{RK}, we
identify $A_h$ as the source emitting a Higgs scalar of mass $m$.
We can also determine the active vertices of the reformulated theory. For instance,
\bqa
\diag{010}{1.4}{-.3} &=& -i{m^2\over v^2}\,T_{knpr}\,x_k\,x_n\,x_p\,x_r = -3i{m^2\over v^2}\;\;,\nl
\diag{011}{1.4}{-.3} &=& \diag{012}{1.4}{-.3} = -3i{m^2\over v}\;\;,   \label{hself}
\eqa
and in a similar way we find
\bq
\diag{015}{1.4}{-.3} = -i{m^2\over v^2}\left(\wph1+2\de_{jk}\right)\;\;\;,\;\;\;
\diag{013}{1.4}{-.3} = -i{m^2\over v^2}\;\;\;,\;\;\;
\diag{014}{1.4}{-.3} = -i{m^2\over v}\;\;.\label{scalarvertices}
\eq
A few observations are important at this point.
In the first place, the identity 
(\ref{counter1}) is precisely that, an {\em identity\/}. Therefore we must use 
\[
\mbox{either}\;\;\;\diag{016}{1.}{-.2}\;\;\;\;\mbox{or}\;\;\;\;\diag{017}{1.2}{-.3}\;\;\;\mbox{but {\em not\/} both}\;\;.
\]
This counter-intuitive-seeming prescription is the basis for the diagrammatic description of
the Higgs mechanism.
  Not taking \eqn{counter1} as an actual physical identity leads to incorrect handling
of the symmetry factors, and possible mis-counting of diagrams.
For example, iterating one leg of the SDe (\ref{counter1}) would give
\bq
\diag{003}{.8}{-.1}\;\; =\;\;  \diag{004}{1.}{-.5}\;\;=\;\;\diag{018}{1.1}{-.4}\;\;,
\eq
with the incorrect implication that $\suml_n\tau_n^2 = v^2/3$. Only if we iterate all legs,
\bq
\diag{003}{.8}{-.1} \;\;=\;\; \diag{019}{1.2}{-.5}\;\;,
\eq
do we again find the correct result. 

Secondly, if we treat the tadpoles diagrammatically we have to admit that they are zero modes, {\it i.e.\/}
they are constant. If we let $\tau_n$ depend on position, the tadpoles act as sources of momentum that will
needlessly tangle our treatment. We therefore take $\kt{x}$ to be a {\em constant\/} unit  vector. 

Finally, the very {\em idea\/} of SSB is that $\kt{x}$ is also a {\em random\/} unit vector,
outside of our control: we are therefore forbidden from making any further assumptions on $\kt{x}$. Of course,
a good model ought to yield physics that is, as much as possible,  independent of $\kt{x}$.

\subsection{A Goldstone infrared problem\label{IRprob}}
In the foregoing we have derived a consistent set of Feynman rules containing one massive scalar and at least one
massless Goldstone scalar. Once we try to truncate the massless internal lines, however, we run into problems
for $N_t>2$, since
it is not clear how we can disentangle the various massless propagators so as to identify the various Goldstones. Worse,
the model contains infrared phenomena.  Let us imagine a process in which a
massless Goldstone is emitted with momentum $p$:
\bq
\calm_0(p) =\;\diag{192}{1.6}{-.4}\;\;.
\eq
The (differential) cross section is given by
\bq
d\si_0 \sim |\calm_0(p)|^2\,d^4p\,\de(p^2)\tha(p^0)\;\;.
\eq
We may let the Goldstone go very slightly off-shell and decay into three. Assuming that $\calm_0(p)$ does not depend
too drastically on $p$, we can write this as
\bq
\calm_1 = \diag{193}{2}{-.4} \approx
\calm_0(q_1+q_2+q_3)\,{1\over(q_1+q_2+q_3)^2}\,{m^2\,k\over v^2}\;\;,\label{threegold}
\eq
where $k=3$ if the three Goldstones are identical, otherwise 1. The cross section, which is dominated by the diagram of
\eqn{threegold}, can then be written as
\bqa
d\si_1 &\sim& |\calm_1|^2\,\prodl_{j=1}^3d^4q_j\,\de(q_j^2)\,\tha(q_j^0)\nl
&&\hspace{1cm}\times\;\de^4(q_1+q_2+q_3-p)\,d^4p\,\de(p^2-u)\,du\nl
&\approx& |\calm_0(p)|^2 d^4p\, \de(p^2-u)\,\tha(p^0)\;{m^4\,k\over2v^4}\,{\pi^2\over8}\,{du\over u}
\approx d\si_0\;{m^4\,k\over2v^4}\,{\pi^2\over8}\,{du\over u} \;\;,\label{IRD}
\eqa
where we have used the result \cite{RK}
\bq
\int\left(\prodl_{j=1}^3d^4q_j\,\de(q_j^2)\,\tha(q_j^0)\right)\,\de^4(q_1+q_2+q_3-p) = {\pi^2\over 8}p^2\;\;.
\eq
We see that the $u$ integral in \eqn{IRD} diverges for $u\to0$: an infrared divergence.\footnote{In a good theory such as
QED, the actual IR divergences are cancelled by corresponding IR divergences from 
virtual corrections. Even if this happens here, the fact remains that for very small $u$
a three-body-decaying Goldstone wins out over a non-decaying one.}
Physically, this means that
an external massless Goldstone in any process will unavoidably emerge as a cloud of low-energy Goldstones, all moving
collinearly with the speed of light. This makes the LSZ truncation for such particles extremely problematic.
It would seem that if we want to end up with a model in which particles can be unambiguously identified,
massless Goldstone bosons are to be avoided.\footnote{One might also worry about internal lines carrying
vanishing momentum, for instance in the vacuum diagram 
$\diag{194}{1}{-.1}$. However, some reflection shows that such zero-momentum lines can only be the massive Higgs propagators.}

\section{The Abelian Higgs model}
\subsection{Feynman rules and current conservation}
We may couple the tachyons to a massless vector (called `photon' for now), with a propagator given by
\bq
\diag{020}{1.7}{-.2} = -{i\over q^2}k^{\al\be}\;\;\;,\;\;\;k^{\al\be} = g^{\al\be} + F^\al q^\be + q^\al F^\be + G\, q^\al q^\be\;\;.
\label{photonprop}
\eq
The quantities $F^\mu$ and $G$ are related to the gauge choice: in this model, they are quite immaterial as we shall see.
We shall assume two interaction vertices between photon and tachyons:\
\bq
\diag{021}{1.6}{-.5} = f^n_k\,(p-q)^\mu\;\;\;,\;\;\;
\diag{022}{1.4}{-.5} = i\,t^n_k\,g^{\al\be}\;\;.
\eq
Because the tachyons are bosons, the real matrices $f$ and $t$ must of necessity be {\em antisymmetric\/} and {\em symmetric}, 
respectively.\footnote{If current conservation (see further on) is to have 
any chance at all, the combination $(p-q)^\mu$ is unavoidable, as is therefore the antisymmetry of $f$.}\\

Since the photon is massless, its {\em observable\/} polarization must be purely transverse 
in {\em any\/} Lorentz frame. That is only
possible if its current is conserved \cite{RK}. As before, we shall give the following steps in some detail since
we will employ them again later on. Let $M(q)^\mu$ be
the complete set of connected diagrams that emit or absorb a photon with momentum $q$ (not necessarily on-shell):
\bq
M(q)^\mu = \diag{023}{1.7}{-.2}\;\;.
\eq
The requirement of current conservation means that
\bq
M(q)^\mu \,q_\mu = \diag{024}{1.7}{-.2}\; \must\; 0\;\;.\label{curcon}
\eq
The symbol $\must$ means that we {\em demand\/} the zero result: we have to construct our theory so as to 
arrange it.
The `handlebar' denotes multiplication with the momentum, considered {\em outgoing}. We have to prove \eqn{curcon}
in full generality, and we shall do this using the SDe. Let us denote by a lightly shaded {\em semi-connected\/}
blob a complete set of diagrams
that are not necessarily connected but do not contain vacuum bubbles. Now consider
\bq
\diag{025}{2}{-.6} = {i\over p^2+m^2/2}\,f^n_k\,(p-q)^\mu\,{i\over q^2+m^2/2}\;\;,
\eq
where we have left out the expression for the semi-connected blob. Applying the handlebar gives us
\bqa
\diag{026}{1.7}{-.65} &=& {i\over p^2+m^2/2}\,f^n_k\,(p-q\cdot -p -q)\,{i\over q^2+m^2/2}\nl
&=& {i\over p^2+m^2/2}\,f^n_k\,i\;\;+\;\;i\,f^k_n\,{i\over q^2+m^2/2}\;\;.\label{handlebar1}
\eqa
Note that this works as long as the mass terms (the $+m^2/2$) are the same in both tachyon propagators: their sign is
irrelevant.
We can represent \eqn{handlebar1} diagrammatically by auxiliary Feynman rules, as follows:
\bq
\diag{027}{1.6}{-.5} = \diag{028}{1.6}{-.5}\;\;\;,\;\;\;\diag{029}{1.2}{-.3} = f^n_k\;\;,\;\;\diag{030}{.8}{-.1} = i\;\;.
\label{auxifirst}
\eq
It is important to note that, unless we specify them explicitly, equivalent lines entering the semiconnected
blob have to be symmetrized over, and therefore \eqn{auxifirst} includes {\em both\/} terms of \eqn{handlebar1}.
In addition, if a slashed line happens to be an external one, it will not survive LSZ truncation since it has no pole:
therefore such diagrams can be neglected.\footnote{This is because we consider S-matrix elements
rather than Green's functions.}
Next, we consider the following tree-level amplitude:\footnote{Written like this, the process is kinematically
impossible. By moving tachyon $n$, say, to the initial state (replacing $p$ by $-p$) we can repair this. The conclusion remains the same.} 
\bq
\diag{031}{2}{-.7}\; =\; \diag{032}{1.6}{-.7} +  \diag{033}{1.6}{-.7} +  \diag{034}{1.6}{-.7}\;\;\must\;\;0\;\;.\label{mustseagull}
\eq
 Applying the
Feynman rules we find
\bq
-i\left(\vph2(f^2)^n_k + t^n_k\right)( p + q + r)^\al \must 0\;\;,
\eq
so that we require
\bq
t^n_k \must - 2\, f^n_\ell f^\ell_k = 2\,f^n_\ell k^k_\ell\;\;.
\eq
We then have
\bq
\diag{035}{1.6}{-.5} \;=\;  -\;\diag{036}{1.6}{-.6}\;\;.
\eq
The proof of general current conservation proceeds as follows, where we leave the semi-connected blobs on the
right to be understood. The SDe reads
\bq
\diag{037}{1.4}{-.2} = \diag{038}{1.4}{-.3} + \diag{039}{1.3}{-.3}\;\;.
\eq
We now apply the handlebar. Since slashed external lines do not contribute, we can SDe-iterate the
slashed propagators by letting them enter a vertex, so that
\bqa
\diag{040}{1.4}{-.3} &=& - \;\diag{041}{1.4}{-.35}\;\;,\nl
\diag{042}{1.3}{-.3} &=& + \;\diag{041}{1.4}{-.35}\;+\;\diag{043}{1.6}{-.45}\;+\;\diag{044}{1.6}{-.45}\;\;.
\eqa
Let $n$, $k$ and $r$ be distinct tachyon labels. Then, dropping unimportant overall factors (denoted by $\sim$), we have
\bqa
\diag{043}{1.6}{-.45} &=& \diag{045}{1.6}{-.45} + \diag{046}{1.6}{-.45} \sim (f^3)^n_k + (f^3)^k_n = 0\;\;,\\
\diag{044}{1.6}{-.45} &=& \diag{047}{1.6}{-.45} + \diag{048}{1.6}{-.45} + \diag{049}{1.6}{-.45} + \diag{050}{1.6}{-.45}\nl
&\sim& \left({3\over3!}\,f^k_n + {1\over2!}\,f^n_k\right) + 
\suml_{r\ne k,n}\left({1\over2!}\,f^n_k + {1\over2!}\,f^k_n\right) = 0\;\;.  \label{antkaz}
\eqa
This finishes the proof of current conservation, \eqn{curcon}: the only condition on $f$ is that it be antisymmetric. 
The presence of tadpoles does not change the argument, since the semi-connected blob may contain tadpoles at will. We
simply must adhere to the convention of
\bq
\mbox{never using}\;\;\;\diag{114}{1.3}{-.2}\;\;\;\mbox{but always}\;\;\;\diag{115}{1.3}{-.2}\;\;.\label{tadpolerule}
\eq
That is, in the SDe iteration the tadpole does not count as a vertex.\\

\subsection{Dressed propagators and masquerading}
Next, we turn to the dressed propagators, the analogues of $\Pi_{nk}$ of the previous section.
We encounter two additional passive vertices, in addition to that of \eqn{passive1}: introducing another vector in tl-space,
\bq
\;\;e_n =  x_\ell\,f^\ell_n\;\;,\;\;\kt{e} = f\kt{x}\;\;,\;\;\bk{e}{e} \equiv e^2\;\;,\;\;\bk{e}{x}=0\;\;,
\eq
we  have
\bqa
\diag{051}{1.7}{-.2} &=& \tau_\ell\, f^\ell_n\,q^\al\;=\;v\,e_n\,q^\al\;\;,\nl
\diag{052}{1.7}{-.2} &=& {1\over2!}\,\tau_k\,\tau_n\,t^n_k\,g^{\al\be}\;
=\; iv^2\,\bk{e}{e}\, g^{\al\be}\;=\;i\,e^2v^2\,g^{\al\be}\;\;.
\label{passive2}
\eqa
There are now three types of dressed propagators, with the momentum $q$ assumed to be moving from left to 
right\footnote{$\Pi$ and $\Om$ are even in $q$, but $\Psi$ is {\em odd\/} in $q$.}:
\bqa
\Pi _{nk}&=& \diag{061}{1.6}{-.1} = \diag{007}{.8}{0} + \diag{008}{1.6}{-.1} + \diag{053}{1.8}{-.2}\;\;,\nl
\Psi_k &=& \diag{054}{1.5}{-.2} = \diag{055}{1.6}{-.2} + \diag{056}{1.6}{-.2}\;\;,\nl
\Om &=& \diag{060}{1.4}{-.15} = \diag{057}{.65}{-.05}\; + \;\diag{058}{1.6}{-.15}\; +\; \diag{059}{1.6}{-.16}\;\;.
\eqa
Following the steps described in section \ref{getgold} we can write\footnote{For typographical reasons we shall always write 
Lorentz indices as upper indices, and the Minkowski metric is understood.}
\bqa
\Pi_{nk}\,q^2 &=& i\de_{nk} + m^2x_n\, x_\ell\,\Pi_{\ell k} + iv\,e_n\, q^\mu\Psi^\mu_k\;\;,\nl
\Psi^\al_k\,q^2 &=& m^2\,\Psi^\al_\ell\, x_\ell\,x_k - iv\,\Om^{\al\mu}q^\mu\, e_k\;\;,\nl
\Om^{\al\be}\,q^2 &=& -ik^{\al\be} + e^2v^2\,\Om^{\al\mu}k^{\mu\be}
                                  - iv\,\Psi^\al_\ell\, e_\ell\,q^\mu k^{\mu\be}\;\;,
\eqa
so that
\bq
 x_\ell\,\Pi_{\ell k} = {i\,x_k\over q^2-m^2}\;\;,\;\;\Psi^\al_\ell\, x_\ell = 0\;\;.                            
\eq
Introducing an orthonormal basis in tl-space (and ignoring the possible interpretational problem raised in section \ref{IRprob}):
\bq
\kt{x}\;\;,\;\;{1\over e}\kt{e}\;\;,\;\;\kt{y^j}\;\;\;(j=3,\ldots,N_t)\;\;\;\Rightarrow\;\;\;
\de_{nk}-x_nx_k = {1\over e^2}e_ne_k + (y^j)_n(y^j)_k\;\;,
\eq
we arrive at the forms
\bqa
\Pi_{nk} &=& {i\over q^2-m^2}x_nx_k + {i\over q^2}\left({1\over e^2}e_ne_k + y^j_n\,y^j_k\right)
+ {v^2\over q^4}\,e_n\,q^\la\,\Om^{\la\si}\,q^\si\,e_k\;\;,\nl
\Psi^\al_k &=& -{iv\over q^2}\,\Om^{\al\la}q^\la\, e_k\;\;,\label{AHprops1}\\
\Om^{\al\be}\,q^2 &=& -ik^{\al\be} + M^2\,\Om^{\al\mu}\,T^{\mu\nu}\,k^{\nu\be}\;\;,\label{AHprops2}
\eqa
with
\bq
M = ev\;\;,\;\;T^{\al\be} = g^{\al\be} - L^{\al\be}\;\;,\;\;L^{\al\be} = {q^\al q^\be/ q^2}\;\;.
\eq
We cannot solve \eqn{AHprops2} for $\Om$, but as it will turn out that does not matter.
Another point worthy of note follows from the expression for $\Pi_{nk}$. As we can see from its
diagrammatic SDe, the result for $x_\ell\Pi_{\ell k}$ is the same as in \eqn{higgsresult}.
This is because the scalar-vector
mixing tadpole contains a vector $\kt{e}=f\kt{x}$, which is orthogonal to $\kt{x}$.
Introducing several vector particles will lead to different vectors $\kt{e}$, but these are also orthogonal
to $\kt{x}$ (see later on).
Thus, the propagator $\Pi_{nk}$ will always contain the term $ix_nx_k/(q^2-m^2)$, in other words
{\em there will always be a distinct Higgs particle}; this is the original assertion made in \cite {Higgs}.\\

A final ingredient is the following. Again using active vertices, we have two SDe's:
\bqa
\diag{062}{1.2}{-.15} &=& \diag{063}{1.2}{-.15} + \diag{064}{1.2}{-.15} + \diag{065}{1.3}{-.15}\;\;,\nl
\diag{066}{1.2}{-.15} &=& \diag{067}{1.2}{-.15} + \diag{068}{1.2}{-.15} + \diag{069}{1.3}{-.15}\;\;.
\eqa
Current conservation can now be expressed by the handlebar diagrams:
\bqa
0 \;=\; \diag{070}{1.2}{-.15} &=& \diag{071}{1.2}{-.15} - \diag{072}{1.4}{-.1} + \diag{073}{1.3}{-.1} 
+ \diag{074}{1.4}{-.1}\;\;,\label{masque}\\
\diag{074}{1.4}{-.1} &=& \diag{075}{1.4}{-.1} +  \diag{072}{1.4}{-.1} + \diag{076}{1.7}{-.1}\;\;.
\eqa
Everything except the active-vertex diagrams 
cancels.\footnote{To appreciate this, it may help to recast diagrams:
\[\diag{073}{1.}{-.1}  + \diag{076}{1.4}{-.1} = \diag{182}{1.1}{-.4} + \diag{183}{1.1}{-.4} = 0\;\;\;,\]
where we can recognize simply a special case of \eqn{antkaz}. Notice the importance of rule (\ref{tadpolerule})!}
 Denoting by $A_\g^\mu$ the amplitude emitting the photon and by
$A_n$ the one emitting tachyon $n$, we see that current conservation implies
\bq
\diag{071}{1.2}{-.15} + \diag{075}{1.5}{-.1} = 0\;\;\;\Rightarrow\;\;\;A_n\,e_n = {i\over v}A_\g^\mu\,q^\mu\;\;, \label{masq}
\eq
again with the momentum $q$ counted outgoing. In this way a combination of tachyonic tadpoles can 
{\em masquerade\/} as a handlebarred vector boson.

\subsection{Amplitudes}
We are now ready to consider an amplitude, as in the previous section:
\bqa
\calm &=& \diag{077}{2.3}{-.1} + \diag{078}{2.3}{-.1} + \diag{079}{2.3}{-.1} + \diag{080}{2.3}{-.1} \nl
&=& A_n\,\Pi_{nk}\,B_k +  A_\g^\al\,\Psi^\al_k\,B_k 
+ A_n(-\Psi^\be_n)B_\g^\be + A_\g^\al\,\Om^{\al\be}B_\g^\be\;\;.
\eqa
Using the results (\ref{AHprops1}) and \eqn{masq} we can write this as
\bq
\calm = A_h\,R_h(q)\,B_h + A_j\,R_0(q)\,B_j + A_\g^\al\,R_M^{\al\be}(q)\,B_\g^\be\;\;,
\eq
where
\bq
R_M^{\al\be}(q) = {i\over M^2}L^{\al\be} +  D^{\al\be}\;\;\;,\;\;\;D^{\al\be} = T^{\al\mu}\Om^{\mu\nu}T^{\nu\be}\;\;.
\eq
Since $T^{\al\mu}k^{\mu\nu}T^{\nu\be} = T^{\al\be}$,  \eqn{AHprops2} implies
\bq
D^{\al\be}\,q^2 = -i\,T^{\al\be} + M^2\,D^{\al\be}\;\;,
\eq
and so we find for the vector propagator
\bq
R_M^{\al\be}(q) = {i\over M^2}L^{\al\be} - {i\over q^2-M^2}T^{\al\be}
= {i\over q^2-M^2}\left(-g^{\al\be} + {q^\al q^\be\over M^2}\right)\;\;,
\eq
the {\em unitary-gauge\/} propagator. As announced, the quantities $F^\mu$ and $G$ of \eqn{photonprop} drop out
because $q^\al T^{\al\be}=0$. Upon LSZ truncation, the massive vector has precisely its three physical
polarization degrees of freedom, including the longitudinal one. 

\subsection{Vertices\label{ahvertices}}
The vertices of \eqn{scalarvertices} are unchanged, but we have to find the scalar-vector interactions. We see
immediately that
\bq
\diag{081}{1.4}{-.3} \sim \bak{x}{f}{x} = 0\;\;\;,\;\;\;\diag{082}{1.4}{-.4} \sim \bak{x}{f}{y^j} \sim \bk{e}{y^j} = 0\;\;.
\eq
The two-photon vertices\footnote{For reasons lost in the mists of time, these are also called
{\em seagull vertices}.} require more care. First, we find
\bq
\diag{083}{1.6}{-.5} = -2i\bak{x}{t}{x}\,g^{\al\be} = 2i\,e^2\,g^{\al\be} = 2i{M^2\over v^2}g^{\al\be}\;\;,\;\;
\diag{084}{1.4}{-.5} = 2i{M^2\over v}g^{\al\be}\;\;.
\eq
However,
\bq
\diag{085}{1.6}{-.5} \sim \bak{x}{t}{y^j} \sim \bak{e}{f}{y^j}\;\;.
\eq
If we want this to vanish we need $f\kt{e} \sim \kt{x}$. As can be seen from appendix \ref{antihermitian}, that is
only possible if $\kt{x}$ is an eigenvector of $f^2$: but $\kt{x}$ is random! The only reasonable
solution is to require that $\kt{x}$ is {\em always\/} an eigenvector, {\it{i.e.}\/} $f^2$ must be $-e^2$ times the
unit matrix, and the number of tachyons must be even. In that case, the vectors $\kt{y^j}$ can be chosen
such that
\bq
f\kt{y^{2s-1}} = e\kt{y^{2s}}\;\;,\;\;f\kt{y^{2s}} = -e\kt{y^{2s-1}}\;\;,\;\;s=2,\ldots,N_t/2\;\;,\label{ychoice}
\eq
so that they can be interpreted as the real and imaginary part of charged scalars, with
\bq
\diag{086}{2.2}{-.5} = e\,(p-q)^\mu\;\;\;,\;\;\;\diag{087}{1.6}{-.5} = 2ie^2\,g^{\al\be}\;\;.
\eq
If $f^2$ is not proportional to unity, we have a model that is consistent and unitary, but lacks an interpretation
in terms of simple charged scalars. Admittedly, since the physics in that case depends on $\kt{x}$, it is not a 
`good' model in the sense discussed above.

For $N_t=2$, the `original' AH model, the Feynman rules found are
precisely those of the sector of the Standard Electroweak Model that contains only $Z^0$ and Higgs particles.
Since the longitudinal polarization vector of a $Z^0$ with momentum $p$ has the form
$\eps_L^\mu = p^\mu/M - \order{M/p^0}$, the application of \eqn{masq} shows that the longitudinal degree of freedom
is essentially scalar tachyons in masquerade,\footnote{The fact that longitudinally polarized vector bosons may be
replaced by appropriate (combinations of) scalars is referred to as the {\em Equivalence Theorem}.\cite{HV}}
 and so unitarity of this sector is proven.

\section{Interacting vectors: general structure}
\subsection{Feynman rules and current conservation}
We now turn to models containing several massless vectors (labelled by $w,x,y,\ldots$), 
that may be interacting with each other. It becomes necessary to postulate their propagator in 
more detail:\footnote{We do not have to `derive' this propagator by `gauge fixing' in some action,
since we do not use actions. The only  requirement is that our resulting theory be current-conserving: recall footnote 2.} 
\bq
\diag{088}{1.6}{-.2} = R_n^{\al\be}(q) = -i{k^{\al\be}(q)\over q^2}\;\;,\;\;
k^{\al\be}(q) = g^{\al\be} - {q^\al n^\be\over(q\cdot n)} - {n^\al q^\be\over(q\cdot n)} + {n^2\,q^\al q^\be\over(q\cdot n)^2}\;\;,
\label{axialprop}
\eq
where $n^\al$ is a fixed vector. This is the {\em axial gauge}, with $n^\al k^{\al\be} = 0$.
We assume a three-vector coupling, defined by
\bq
\diag{089}{2}{-.8} = h^{wxy}\,Y(p,\al;q,\be;-p-q,\la)\;\;,
\eq
with\footnote{This is just the tachyon-tachyon-vector vertex, adapted to three vectors, that have to be treated on the
same footing (being bosons). It is therefore essentially the simplest possible choice.}
\bq
Y(p,\al;q,\be;r,\la) = (p-q)^\la g^{\al\be} + (q-r)^\al g^{\be\la} + (r-p)^\be g^{\la\al}\;\;.
\eq
Since the vectors are bosons, the real numbers $h^{wxy}$ are necessarily totally antisymmetric. We also propose a
four-vector interaction:
\bq
\diag{090}{2}{-.8} = i\,X(w,\al;x,\be;y,\mu;z,\nu)\;\;,
\eq
with
\bqa
&&X(w,\al;x,\be;y,\mu;z,\nu) = g^{\al\be}g^{\mu\nu}\{wyzx\}
+ g^{\al\mu}g^{\be\nu}\{wxzy\} + g^{\al\nu}g^{\be\mu}\{wxyz\}\;\;,\nl
&&\{wxyz\} = h^{wxt}h^{yzt} + h^{wyt}h^{xzt}\;\;.\label{fourvectors}
\eqa
We need to work out what the handlebar does here:
\bq
\diag{091}{2}{-.6} \sim R_n^{\mu\al}(p)\,h^{wxy}\,Y(p,\al;q,\be;-p-q,\la)(-p-q)^\la\,R_n^{\be\nu}(q)\;\;.
\eq
Using the two results
\bqa
&&Y(p,\al;q,\be;-p-q,\la)(-p-q)^\la = \left(p^\al p^\be - p^2g^{\al\be}\right) - \left(q^\al q^\be - q^2g^{\al\be}\right)\;\;,\nl
&&R_n^{\mu\al}(p) \left(p^\al p^\be - p^2g^{\al\be}\right) = i\left(g^{\mu\al}-{p^\mu n^\al\over(p\cdot n)}\right)\,g^{\al\be}\;\;,
\eqa
we find immediately the rule\footnote{There is a subtlety if one of the vectors is external. Appendix \ref{externalcon}
shows that this does not lead to problems.}
\bq
\diag{092}{1.7}{-.65} = \diag{093}{1.7}{-.65} \;\;,\;\;
\diag{094}{1.3}{-.5} = h^{wxy}g^{\al\be}\;\;,\;\;\diag{095}{1.3}{-.1} = i\,g^{\al\be}\;\;.\label{vec3rule}
\eq
In addition, we have
\bq
\diag{096}{1.7}{-.65} = -\;\diag{097}{1.7}{-.65}\;\;,
\eq
which is the {\it raison d'\^etre\/} of the cumbersome expression (\ref{fourvectors}).
For the interactions with the tachyons, we simply generalize those of the AH model: 
\bq
\diag{107}{1.6}{-.5} = (f^w)^n_k\,(p-q)^\mu\;\;\;,\;\;\;
\diag{108}{1.4}{-.5} = i\,(t^{wx})^n_k\,g^{\al\be}\;\;,
\eq
with again $f^w$ antisymmetric and $t^{wx}=t^{xw}$ symmetric in tl-space. 
By examing the tree-level handlebar condition
\bq
\diag{109}{1.8}{-.4} = \diag{110}{1.8}{-.4}  +  \diag{111}{1.8}{-.4}   +  \diag{112}{1.8}{-.4}   +  \diag{113}{1.6}{-.4} \must 0
\eq
we find the analogue to \eqn{curcon}:
\bqa
[f^w,f^x] \must h^{wxy}\,f^y\;\;\;,\;\;\;\{f^w,f^x\} \must -t^{wx} \label{falgebra}\;\;.
\eqa
With these rules and tools we can prove current conservation for this type of model. Again leaving out the
semi-connected blobs, we write the SDe as
\bq
\diag{037}{1.4}{-.2} = \diag{038}{1.4}{-.3} + \diag{039}{1.3}{-.3} + \diag{098}{1.3}{-.3}  + \diag{099}{1.3}{-.35}\;\;,
\eq
and we have
\bqa
\diag{100}{1.3}{-.3} &=& - \;\diag{041}{1.4}{-.3}\; -\; \diag{101}{1.4}{-.4}\;\;,\nl
\diag{042}{1.3}{-.3} &=& + \;\diag{041}{1.4}{-.35}\;+\;\diag{043}{1.6}{-.45}\;+\;\diag{044}{1.6}{-.45}\;\;,\nl
\diag{102}{1.3}{-.3} &=& -\;\diag{103}{1.2}{-.4}\;\;,\nl
\diag{105}{1.2}{-.3}&=&+ \diag{103}{1.2}{-.4} +  \diag{104}{1.2}{-.7} + \diag{101}{1.4}{-.4} + \diag{106}{1.4}{-.4}\;\;.
\eqa
By methods similar to those leading to \eqn{antkaz} we can straightforwardly show that
\bq
\diag{104}{1.2}{-.7} = 0\;\;\;,\;\;\;\diag{043}{1.6}{-.45} + \diag{106}{1.4}{-.4} = 0\;\;,
\eq
admittedly with extensive use of \eqn{falgebra}. We see that the currents for the massless vectors
are strictly conserved.

\subsection{Dressed propagators, masquerading and the amplitude}
The passive vertices of the model are generalizations of those of the AH model: in addition to \eqn{passive1} we have
\bqa
\diag{119}{1.7}{-.2} &=& \tau_\ell\, (f^w)^\ell_n\,q^\al\;=\;v\,(e^w)_n\,q^\al\;\;,\nl
\diag{120}{1.7}{-.2} &=& {1\over2!}\,\tau_n\,\tau_k\,(t^{wx})^n_k\,g^{\al\be}\;=\;i\,v^2\,\bk{e^w}{e^x}\,g^{\al\be}\;\;,
\label{passive3}
\eqa
with $\kt{e^w} = f^w\kt{x}$, or more explicitly $(e^w)_n = (f^w)^k_nx_k$.
The SDe's for the dressed propagators:
\bqa
\Pi _{nk}&=& \diag{061}{1.6}{-.1} = \diag{007}{.8}{0} + \diag{008}{1.6}{-.1} + \diag{053}{1.8}{-.2}\;\;,\nl
\Psi_{wk} &=& \diag{121}{1.5}{-.2} = \diag{055}{1.6}{-.2} + \diag{056}{1.6}{-.2}\;\;,\nl
\Om_{wx} &=& \diag{122}{1.4}{-.15} = \diag{057}{.65}{-.05}\; + \diag{058}{1.6}{-.15} + \diag{059}{1.6}{-.1}\;\;,
\eqa
can be treated in the same way as before, to yield
\bqa
\Pi_{nk} &=& {i\over q^2-m^2}x_nx_k + {i\over q^2}(\de_{nk}-x_nx_k) + 
                     {v^2\over q^4}(e^w)_n\,q^\mu\Om_{wx}^{\mu\nu}q^\nu(e^x)_k\;\;,\nl
\Psi_{wk}^\al &=& -{iv\over q^2}\Om_{wy}^{\al\mu}q^\mu(e^y)_k\;\;,\nl
\Om_{wx}^{\al\be}\, q^2&=& -ik^{\al\be}\de_{wx} + v^2\,\Om_{wy}^{\al\mu}T^{\mu\nu}k^{\nu\be}\bk{e^y}{e^x}\;\;.
\eqa

If we denote by $A_w^\mu$ the active-vertex source emitting vector $w$, we can perform the same steps as for \eqn{masq}
and find the masquerading identities
\bq
A_n(e^w)_n = {i\over v}A_w^\mu q^\mu\;\;,\label{masqn}
\eq
that we shall use extensively.
We do have to note the appearance of an extra term on the right-hand side in \eqn{masque}:
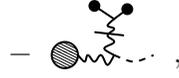
\[  -\; \;\diag{123}{1.4}{-.1}\;\;,\]
but since the tadpoles carry no momentum this diagram vanishes by itself.

The amplitude, with dressed propagators running between active vertices, is also treated as before, and we
find
\bq
\calm = A_hR_h(q)B_h + A_n{1\over q^2}(\de_{nk}-x_nx_k)B_k +
A_w^\al D_{wx}^{\al\be}B_x^\be\;\;,
\eq
with
\bq
D_{wx}^{\al\be} =T^{\al\mu}\Om_{wx}^{\mu\nu}T^{\nu\be} =  -iT^{\al\be}\,K_{wx}\;\;\;,\;\;\;
K_{wx}\,q^2 = \de_{wx} + v^2\,K_{wy}\,\bk{e^y}{e^x}\;\;.
\label{generalD}
\eq
Further evaluation  depends on the details of the model: in the following we shall examine 
several such models. It may be interesting to point out that the choice of the axial propagator of
\eqn{axialprop}
is necessary to make \eqn{vec3rule} work, but that is the {\em only\/} place in the whole discussion where
it plays a r\^ole: in particular the propagators $D^{\al\be}$ are independent of the gauge choice.\footnote{And the
axial-gauge vector $n^\mu$ disappears completely from our discussion, as it {\em ought to}.}

\section{Interacting vectors: example models} 
\subsection{The Apollo model\label{apollo}}
The minimum number $N_v$ of vector fields that can be self-interacting is 3, in which case $h^{wxy}$ must be
proportional to the Levi-Civita symbol $\vreps^{wxy}$. Since the $f$ matrices are antisymmetric, the minimum
number $N_t$ of tachyons is also 3.\footnote{For $N_t=2$ all the $f$ matrices would commute.}
Therefore the Apollo model (first discussed in \cite{GG}), with $N_t=N_v=3$, is 
the `smallest' model with vector self-interactions. Let us define the $3\times3$ matrices $f^w$ for this model by
\bq
(f^w)^n_k = -e\,O^w_y\,\vreps^{ynk}\;\;,
\eq
where $O$ is an arbitrary but fixed orthogonal $3\times3$ matrix. It can be checked that, indeed, 
\bq
[f^w,f^x] = h^{wxy}f^y\;\;\;,\;\;\;h^{wxy} = e\,\vreps^{wxy}\;\;.
\eq
Since $\bk{e^w}{x}=0$, the three vectors $\kt{e^w}$,
\bq
(e^w)_k = e\,O^w_y\vreps^{ykn}x_n\;\;,\;\;w=1,2,3\;\;,
\eq
cannot be linearly independent. If we define
\bq
\g_w = O^n_w\,x_n\;\;\;,\;\;\;\bk{\g}{\g} = 1\;\;,
\eq
then we can easily verify that
\bq
\g_w\,\kt{e^w} = 0\;\;\;,\;\;\;\bk{e^w}{e^x} = e^2(\de_{wx} - \g_w\g_x)\;\;.
\eq
\eqn{generalD} now takes the form
\bq
K_{wx}\,q^2 = \de_{wx} + M^2K_{wx} - M^2K_{wy}\g_y\g_x\;\;,
\eq
from which we readily derive
\bq
K_{wx} = {1\over q^2}\g_w\g_x + {1\over q^2-M^2}(\de_{wx}-\g_w\g_x) =
 {1\over q^2}\g_w\g_x + {1\over q^2-M^2}(\ro_w\ro_x +\tau_w\tau_x) \;\;,
\eq
where the three unit vectors $\vec{\g}$, $\vec{\ro}$ and $\vec{\tau}$ form a complete orthonormal set,\footnote{The 
vector $\vec{\tau}$ is not to be confused with the tachyonic tadpoles $\tau_n$.}
with $\g_w\ro_x\tau_y \vreps^{wxy} = 1$.
Defining
\bq
A^\mu_\g = A^\mu_w \g_w\;,\;\;A^\mu_\ro = A^\mu_w \ro_w\;,\;\;A^\mu_\tau = A^\mu_w \tau_w\;,\;\;
\eq
we find immediately that $A_\g^\mu$ is strictly conserved since 
\bq
q^\mu A^\mu_\g = (q^\mu A_w^\mu)\g_w = -iv\,A_n(e^w)_n\g_w = 0\;\;.
\eq
Moreover, again using completeness we can derive, with $\kt{e^r}=r_w\kt{e^w}$:
\bqa
\lefteqn{\hspace{-2cm}(e^\ro)_n(e^\ro)_k + (e^\tau)_n(e^\tau)_k = e^2(\ro_u\ro_y+\tau_u\tau_y+\g_u\g_y)(e^u)_n(e^y)_k}\nl
&=& e^2(e^u)_n(e^u)_k = e^2O^u_aO^u_b\,\vreps^{an\ell}\vreps^{bkr}x_\ell x_r = 
e^2\vreps^{an\ell}\vreps^{akr}x_\ell x_r\nl
&=& e^2(\de_{nk} - x_nx_k)\;\;.
\eqa
The amplitude can therefore be written as
\bq
\calm = A_hR_h(q)B_h + A_\g^\al R_\g^{\al\be}(q)B_\g^\be + \suml_{\si=\ro,\tau}A_\si^\al R_M^{\al\be}(q)B_\si^\be\;\;.
\eq
with
\bq 
R_\g^{\al\be}(q) = -i{g^{\al\be}\over q^2}\;\;;
\eq
and we recognize one Higgs, one massless vector (a `photon') and two massive vectors.
\footnote{In \cite{GG} this was employed to arrive at an `electroweak' model without $Z$ bosons.}\\

We now turn to the vertices: \eqn{hself} still holds. Because of the form of $\bk{e^w}{e^x}$ we see that $\g$ does not couple
to the scalars, while\footnote{We leave the effect of replacing one $h$ by the tadpole, turning the four-point vertex into a three-point one and giving an extra factor $v$, as understood.}
\bq
\diag{124}{1.6}{-.5} = \diag{125}{1.6}{-.5} = 2i{M^2\over v^2}g^{\al\be}\;\;\;,\;\;\;\diag{126}{1.6}{-.5} = 0\;\;.
\eq
The three-vector coupling can be written as
\bq
\diag{127}{2}{-.8} = \g_w\ro_x\tau_y h^{wxy}\,Y(p,\al;q,\be;-p-q,\la) = e\,Y(p,\al;q,\be;-p-q,\la)\;\;.
\eq
For the four-vector coupling we have
\bq
\{aabb\} = \{abab\} = e^2\;\;,\{abba\} = -2e^2\;\;\;,\;\;\;(a,b) = (\g,\ro), (\g,\tau)\;\mbox{or}\;(\ro,\tau)\;\;,
\eq
all other combinations vanishing. Therefore,
\bq
\diag{128}{2}{-.8} = -ie^2\left(2g^{\al\be}g^{\mu\nu} - g^{\al\mu}g^{\be\nu} - g^{\al\nu}g^{\be\mu}\wph\right)\;\;.
\eq

At this point we can start talking about {\em charged\/} vectors, denoted by $W^+$ and $W^-$ (or $+$ and $-$).
We define\footnote{The difference in definition for $A_\pm$ and $B_\pm$ is that $A$ {\em emits}, and $B$ {\em absorbs\/}
the charged vector.}
\bq
A^\mu_\pm = {1\over\sqrt{2}}(A^\mu_\ro\pm iA_\tau^\mu)\;\;,\;\;
B^\mu_\pm = {1\over\sqrt{2}}(B^\mu_\ro\mp iB_\tau^\mu)\;\;.
\eq
Then,
\bq
 A_\ro^\al R_M^{\al\be}(q)B_\ro^\be + A_\tau^\al R_M^{\al\be}(q)B_\tau^\be =  
 A_+^\al R_M^{\al\be}(q)B_+^\be + A_-^\al R_M^{\al\be}(q)B_-^\be
 \eq
 Note that which one of the two terms ($W^+$ or $W^-$) 
 actually survives depends on what happens {\em after\/} the active vertices.
 We therefore have to symmetrize over the lines $\ro,\tau$ and also over the $W^+,W^-$ lines. For once replacing 
 wavy lines by smooth lines for readability, and using $\pm$ for $W^\pm$, we therefore write
 \bqa
&& \diag{131}{2}{-.7} +  \diag{132}{2}{-.7} \;\;=\;\;  \diag{133}{2}{-.7} +  \diag{134}{2}{-.7}\;\;,\nl
&& \diag{135}{1.3}{-.8}\; + \diag{136}{1.3}{-.8}\; + \mbox{($4$ others)} = 
\diag{137}{1.3}{-.8}\; + \diag{138}{1.3}{-.8} + \mbox{($4$ others)}\;\;.
 \eqa
For the $W^+W^-\g\g$ and $W^+W^-$-Higgs couplings we can follow the same argument. This gives us the Feynman rules
 \bqa
\diag{129}{2}{-.8} &=&  ie\,Y(p,\al;q,\be;-p-q,\la)\;\;,\nl
\diag{130}{2}{-.8} &=& -ie^2\left(2g^{\al\be}g^{\mu\nu} - g^{\al\mu}g^{\be\nu} - g^{\al\nu}g^{\be\mu}\wph\right)\;\;,\nl
\diag{139}{2}{-.8} &=& ie^2\left(2g^{\al\be}g^{\mu\nu} - g^{\al\mu}g^{\be\nu} - g^{\al\nu}g^{\be\mu}\wph\right)\;\;,\nl
\diag{140}{1.6}{-.6} &=& 2i{M^2\over v^2}g^{\al\be}\;\;.
\eqa

\subsection{The Higgs-Kibble model}
The HK model arises if we enlarge the tachyon space of the Apollo model to $N_t=4$. The general  form of the
$f^w$ matrices is now given by
\bq
(f^w)^n_k = e\,O^w_z(g^z)^n_k\;\;,\label{fmatrices}
\eq
where the orthogonal matrix $O$ is as in the previous section,\footnote{The three vector bosons are
mixed in an arbitrary way.} and, in block notation
\bq
g^1 = \bmat{cc}$S$&0\\0&$S$\emat\;\;,\;\;g^2 = \bmat{cc}0&$\si_1$\\$-\si_1$&0\emat\;\;,\;\;
g^3 = \bmat{cc}0&$\si_3$\\$-\si_3$&0\emat\;\;,\label{gmatrices}
\eq
where $\si_{1,2,3}$ are the Pauli matrices, and 
$S=i\si_2$.\footnote{A derivation is given in 
appendix \ref{derivegs}.}
The matrices $f$ have the following properties:
\bq
[f^w,f^x] = 2e\,\vreps^{wxy}f^y\;\;,\;\;\{f^w,f^x\}^n_k = -2e^2\,\de^{wx}\,\de^{nk}\;\;.
\eq
Since this implies $\bk{e^w}{e^x} = e^2\de^{wx}$, we find immediately that
\bq
K_{wx} = {\de_{wx}\over q^2-M^2}\;\;.
\eq
Next, from
\bq
\left(\;\wph\kt{e^w}\br{e^w}\;\right)\kt{x}=0\;\;,\;\; \left(\;\wph\kt{e^w}\br{e^w}\;\right)\kt{e^x} = \kt{e^x}
\eq
it follows that
\bq
\de_{nk} - x_nx_k = {1\over e^2}\,(e^w)_n(e^w)_k\;\;.\label{ecomplete}
\eq
We find the amplitude in a straightforward manner:
\bq
\calm = A_hR_h(q)B_h + \suml_wA_w^\al R_M^{\al\be}(q)B_w^\be\;\;,
\eq
so that we have three mass-$M$ vectors and one Higgs scalar. Since the three vectors are all on an equal footing,
not much would be gained by trying to introduce the notion of `charge' in the HK model at this point.\footnote{This
changes once we introduce fermions, see later on.}

\subsection{The electroweak model\label{ewmodel}}
We can extend the HK model in the following way. It is possible to choose, in addition to the matrices $f^j$ ($j=1,2,3$) of
\eqn{fmatrices}, a single matrix $f^0$ that commutes with each $f^j$, that is,
we can add a single vector boson that has no interactions with the other three.\footnote{Only one such matrix $f^0$ 
can be chosen,
see appendix \ref{derivegs}.} Consistently, we can choose $h^{0wx}=0$, and the proof of current conservation
then goes precisely as before. We also define the vector $\kt{e^0} = f^0\kt{x}$. 
This must be a linear combination of the $\kt{e^w}$,
and we write
\bq
\bk{e^w}{e^x} = e^2\de_{wx}\;\;,\;\;\bk{e^0}{e^0} = e'^2\;\;,\;\;\bk{e^w}{e^0} = ee'z_w\;\;,\;\;z_wz_w = 1\;\;.
\eq
Application of \eqn{generalD} then gives, with $M=ev$ and $M'=e'v$,
\bqa
K_{wx}(q^2-M^2) &=& \de_{wx} + MM'K_{w0}z_x\;\;,\nl
K_{w0}(q^2-M'^2) &=& MM'K_{wx}z_x\;\;,\nl
K_{00}(q^2-M'^2) &=& 1 + MM'K_{w0}z_w\;\;.
\eqa
After a little algebra\footnote{Helped by first working out $K_{wx}z_x$ and then $K_{w0}$.} we find
\bqa
K_{wx} &=& {z_wz_x\cth^2 \over q^2-N^2} + {z_wz_x\sth^2\over q^2} + {\de_{wx}-z_wz_x\over q^2-M^2}\;\;,\nl
K_{w0} &=& {z_w\cth\sth\over q^2-N^2} - {z_w\cth\sth\over q^2}\;\;,\nl
K_{00} &=& {\sth^2\over q^2-N^2} + {\cth^2\over q^2}\;\;,
\eqa
where we have introduced
\bq
N^2 = M^2+M'^2\;\;\;,\;\;\;M = \cth N\;\;,\;\;M' = \sth N\;\;\;,\;\;\;\cth^2+\sth^2 = 1\;\;.
\eq
We are therefore naturally led to define
\bq
A_Z^\mu = \cth A^\mu_wz_w + \sth A_0^\mu\;\;\;,\;\;\;A_\g^\mu = -\sth A^\mu_wz_w + \cth A_0^\mu\;\;.
\eq
From \eqn{ecomplete} we derive
\bq
(e^w)_nz_w = {1\over ee'}(e^w)_n\bk{e^w}{e^0} = {e\over e'}(e^0)_n = {\cth\over\sth}(e^0)_n\;\;,
\eq
and we see that $A_\g$ is strictly conserved:
\bq
q^\mu A_\g^\mu \sim -\sth A_n(e^w)_nz_w + \cth A_n(e^0)_n = 0\;\;.
\eq
We can complement the vector $\vec{z}$ by two unit vectors $\vec{r}$ and $\vec{t}$ into an orthonormal set, so that
\bq
\de_{wx} - z_wz_x = r_wr_x + t_wt_x\;\;,
\eq
and define $A_s^\mu = A_w^\mu s_w$ and $\kt{e^s} = s_w\kt{e^w}$ ($s=r,t$), so that we can write
\bq
\de_{nk} - x_nx_k = {1\over e^2}\left(\wph (e^r)_n(e^r)_k + (e^t)_n(e^t)_k + z_w(e^w)_nz_x(e^x)_k\right)\;\;.
\eq
Finally, realizing that
\bq
A_n(e^w)_nz_w = {i\over v}\,q^\mu A_w^\mu z_w = {i\over v}\,q^\mu\left(\cth A^\mu_Z - \sth A_\g^\mu\right) 
= i{\cth\over v}\, q^\mu A_Z^\mu\;\;,
\eq
we find that the amplitude can be written as
\bq
\calm = A_hR_h(q)B_h + \suml_{s=r,t}A_s^\al R_M^{\al\be}(q)B_s^\be
+ A_Z^\al R_N^{\al\be}(q)B_Z^\be + A_\g^\al R_\g^{\al\be}(q)B_\g^\be\;\;.
\eq
We recognize a Higgs scalar, two `$W$' particles of mass $M$,  a $Z$ particle of mass $N$, and a massless photon.

Turning to the vertices, we first note that
\bqa
&&\kt{e^\g} = \cth\kt{e^0} - \sth z_w\kt{e^w} = 0\;\;,\;\;\kt{e^Z} = \cth z_w\kt{e^w} + \sth\kt{e^0} = {1\over\sth}\kt{e^0}\;\;,\nl
&&\bk{e^Z}{e^Z} = {N^2\over v^2}\;\;,\;\;\bk{e^Z}{e^s} \sim z_w\bk{e^w}{e^x}s_x = 0\;\;,\;\;\bk{e^s}{e^s} = {M^2\over v^2}\;\;.
\eqa
Introducing $W^\pm$ as in section \ref{apollo}, the nonzero vector-Higgs couplings are
\bq
\diag{140}{1.6}{-.6} = 2ig^{\al\be}{M^2\over v^2}\;\;\;,\;\;\;\diag{141}{1.6}{-.5} = 2ig^{\al\be}{N^2\over v^2}\;\;.
\eq
The vector self-interactions are simply obtained by using a factor $\cth$ for each $Z$ leg and a factor $-\sth$ for
each $\g$ leg, and remembering that $h^{wxy}$ is now $2e\vreps^{wxy}$ rather than $e\vreps^{wxy}$:
the {\em weak coupling\/} is $g=2e$.

\section{Inclusion of fermions: general structure}
\subsection{Feynman rules and current conservation}
We shall now describe how Dirac fermions can be included in our treatment. We start with massless, chiral fermions, 
that can be left- or right-handed ($L$ or $R$):
\bq
\diag{144}{1}{-.2} = {i\,\om_-\fs{q}\over q^2}\;\;\;,\;\;\;\diag{145}{1}{-.2} = {i\,\om_+\fs{q}\over q^2}\;\;.
\eq
Of these fermions (labelled by $a, b, c,\ldots$) there may be any number.
The chirality projection operators are
\bq
\om_\pm = \half\left(1\pm\g^5\right)\;\;.
\eq
The fermions couple to the vectors and to the tachyons with the following Feynman rules:
\bq
\diag{142}{1.3}{-.2} = iQ^w\g^\mu\;\;\;,\;\;\;\diag{143}{1.3}{-.2} = -iK^n\;\;.
\eq
The objects $Q^w$ and $K^n$ are matrices in fermion-label space (fl-space), and they have to be Hermitian (see 
appendix \ref{cutrule}). Note that $Q^w$ couples $LL$ fermion pairs and $RR$ fermion pairs, while
$K^n$ couples to $LR$ and $RL$ pairs. Splitting fl-space into $L$ and $R$ sectors, we therefore have
\bq
Q^w = \bmat{cc}$Q_L^w$ & 0\\0 & $Q_R^w$\emat\;\;\;,\;\;\;
K^n = \bmat{cc}0 & $\ka_n$\\$\ka_n^\dag$ & 0\emat\;\;.
\eq
Furthermore, it will be handy to distinguish $R$ labels and $L$ labels by dotting the $R$ labels, so that
for instance we have $(Q^w_L)^a_b$, $(Q_R^w)^{\dot{a}}_\db$, and $(\ka_n)^a_\db$. The handlebar rule
with fermions is now seen to be \cite{RK}
\bq
\diag{146}{1.7}{-.5} \;=\; \diag{147}{1.7}{-.5}\;-\;\diag{148}{1.7}{-.5}\;\;\;,\;\;\;
\diag{149}{1.2}{-.3} = \diag{1491}{1.2}{-.3} = iQ^w\;\;,\;\;
\diag{150}{.8}{-.1} = i\;\;.
\eq
This rule is only valid for {\em massless\/} chiral fermions. Further information can be gleaned from the
tree-level four-point amplitudes:
\bqa
\diag{151}{1.5}{-.4} & = & \diag{152}{1.5}{-.5}\;-\;\diag{153}{1.5}{-.5}\;+\;\diag{154}{1.5}{-.5}\;\;\must 0\;\;,\nl
\diag{155}{1.5}{-.4} & = & \diag{156}{1.5}{-.5}\;-\;\diag{157}{1.5}{-.5}\;+\;\diag{158}{1.5}{-.5}\;\;\must 0\;\;,
\label{fermcurcon}
\eqa
from which
\bq
[Q^w,Q^x] \must -ih^{wxy}Q^y\;\;\;,\;\;\;[Q^w,K^n] \must i(f^w)^n_\ell K^\ell\;\;.\label{QKalgebra}
\eq
From the block notation we can write this also as
\bq
[Q^w_J,Q^x_J] \must -ih^{wxy}Q_J^y\;\;(J=L,R)\;\;\;,\;\;\;
Q^w_L\ka_n - \ka_nQ^w_R \must i(f^w)^n_\ell\ka_\ell\;\;.
\eq 
The Jacobi identity $[Q^w,[Q^x,K^n]] - [Q^x,[Q^w,K^n]] = [[Q^w,Q^x],K^n]$ provides a consistency check on
\eqn{QKalgebra}. While the $K^n$ have to be hermitian,
\eqn{QKalgebra} shows that the $\ka_n$ cannot all  be  real.

The equations (\ref{fermcurcon}) also allow us to prove current conservation for our models with fermions in
an almost trivial manner, by including the fermions appropriately in the SDe:
\bqa
\diag{037}{1.4}{-.2} &=& \diag{038}{1.4}{-.3} + \diag{039}{1.3}{-.3} + \diag{098}{1.3}{-.3}  + \diag{099}{1.3}{-.35}
+ \diag{174}{1.3}{-.3}\;\;,\nl
\diag{100}{1.3}{-.3} &=& - \;\diag{041}{1.4}{-.3}\; -\; \diag{101}{1.4}{-.4}\;-\;\diag{175}{1.4}{-.4}\;\;,\nl
\diag{042}{1.3}{-.3} &=& + \;\diag{041}{1.4}{-.35}\;+\;\diag{043}{1.6}{-.45}\;+\;\diag{044}{1.6}{-.45}\;+\;\diag{176}{1.6}{-.45}\;\;,\nl
\diag{102}{1.3}{-.3} &=& -\;\diag{103}{1.2}{-.4}\;\;,\nl
\diag{105}{1.2}{-.3}&=&+ \diag{103}{1.2}{-.4} +  \diag{104}{1.2}{-.7} + \diag{101}{1.4}{-.4} + 
\diag{106}{1.4}{-.4}\;+\;\diag{175}{1.4}{-.4}\;\;,\nl
\diag{177}{1.2}{-.3} &=&+ \diag{178}{1.5}{-.4}\;+\;\diag{179}{1.5}{-.4}\;-\;\diag{180}{1.5}{-.4}\;-\;\diag{181}{1.5}{-.4}\;\;.
\eqa

\subsection{Dressed propagators, amplitudes and vertices}
There are two passive vertices involving fermions:
\bq
\diag{159}{1.7}{-.3} = -i(Y)^a_\db \;\;,\;\;\diag{160}{1.7}{-.3} = -i(Y^\dag)^\da_b\;\;\;,\;\;\;Y = \tau_n\ka_n\;\;.
\eq
The matrices $\ka_n$ are not necessarily square, since we do not have to have equal numbers of $L$ and $R$
fermions.\footnote{As in the Weinberg-Salam model without right-handed neutrinos.}
 This can be remedied by adding an appropriate number of
{\em barren\/} fermions,\footnote{Not to be confused with {\em sterile\/} ones, that have
vanishing $Q$ entries but nonzero $K$ ones.} that have no interactions whatsoever
(zero entries in the $Q^w$ and $K^n$ matrices) and therefore cannot influence the physics. Having done that we can 
employ singular value decomposition; 
that is, we can find unitary matrices $U^a_b$ and $V^\da_\db$ in fl-space such that
\bq
(YY^\dag)^a_b = (U^\dag)^a_cD^c_\dd (D^\dag)^\dd_eU^e_b\;\;,\;\;
(Y^\dag Y)^\da_\db = (V^\dag)^\da_\dc(D^\dag)^\dc_dD^d_\dee V^\dee_\db\;\;,\label{singvaldec}
\eq
with 
\bq
D^a_\db = m_a\,\de_{a\db}\;\;,\;\;(D^\dag)^\da_b = m_a\,\de_{\da b}\;\;\;\;\;\nosum\;\;, \label{formofD}
\eq
where the $m_a$ are nonnegative and real.\footnote{Strictly speaking, singular value decomposition
also works if the  matrix $Y$ is not square. However, in that case $U$ and $V$ have different dimension, and the
matrix $D$ is not diagonal. By employing barren fermions we can use the diagonal form (\ref{formofD}).}
We can then write 
\bq
Y^a_\db = (U^\dag)^a_c D^c_\dd V^\dd_\db\;\;\;,\;\;\;D^a_\db = U^a_cY^c_\dd (V^\dag)^\dd_\db\;\;.
\eq
There are four dressed fermion propagators. We can define
\bq
\Si_{LL} = \diag{161}{1.6}{-.2} = \diag{162}{1}{-.2} + \diag{163}{2.5}{-.3}
\eq
Again letting the momentum $q$ run from left to right, we find
\bq
(\Si_{LL})^a_b\,q^2 = i\om_-\fs{q}\,\de_{ab} + \om_-(YY^\dag)^a_c(\Si_{LL})^c_b\;\;,
\eq
so that
\bq
(U\Si_{LL}U^\dag)^a_b = {i\om_-\fs{q}\over q^2-m_a^2}\de_{ab}\;\;.
\eq
Similarly, we have
\bq
\Si_{RR} =  \diag{164}{1.6}{-.2} \;\;\;\Rightarrow\;\;\;
(V\Si_{RR}V^\dag)^\da_\db = {i\om_+\fs{q}\over q^2-m_a^2}\de_{\da\db}\;\;.
\eq
For one of the mixing propagators,
\bq
\Si_{LR} = \diag{165}{1.6}{-.2} = \diag{166}{2}{-.2}\;\;,
\eq
we find
\bq
(U\Si_{LR}V^\dag)^a_\db = {i\om_-m_a\over q^2-m_a^2}\de_{a\db}\;\;,
\eq
and for the other mixing propagator we finally have
\bq
(V\Si_{RL}U^\dag)^\da_b = {i\om_+m_a\over q^2-m_a^2}\de_{\da b}\;\;.
\eq
A general amplitude involving a fermion line between active vertices has the form
\bqa
\calm &=& \diag{167}{2.3}{-.3}\;+\; \diag{168}{2.3}{-.3}\;+\; \diag{169}{2.3}{-.3}\;+\; \diag{170}{2.3}{-.3}\nl
&=& \Bbar_L\Si_{LL}A_L + \Bbar_R\Si_{RR}A_R +\Bbar_L\Si_{LR}A_R +\Bbar_R\Si_{RL}A_L\;\;.
\eqa
We now introduce unitarily transformed amplitudes:\footnote{This convention is consistent since $A$ is a spinor,
and $B$ is a {\em conjugate\/} spinor.}
\bq
\hat{A}_L = UA_L\;\;,\;\;\hat{\Bbar}_L = \Bbar_L U^\dag\;\;,\;\;\hat{A}_R = VA_R\;\;,\;\;\hat{\Bbar}_R = \Bbar_R V^\dag\;\;.
\eq
The amplitude then becomes (summing over $a,\da$)
\bqa
\calm &=& 
(\hat{\Bbar}_L)_a{i\om_-\fs{q}\over q^2-m_a^2}(\hat{A}_L)^a +
(\hat{\Bbar}_R)_\da{i\om_+\fs{q}\over q^2-m_a^2}(\hat{A}_R)^\da \nl
&&+\; (\hat{\Bbar}_L)_a{i\om_-m_a\over q^2-m_a^2}(\hat{A}_R)^\da +
(\hat{\Bbar}_R)_\da{i\om_+m_a\over q^2-m_a^2}(\hat{A}_L)^a\nl
&=& \Bbar_a {i(\fs{q}+m_a)\over q^2-m_a^2} A_a\;\;,
\eqa
where we have collected the various chiral amplitudes:
\bqa
A &=& \om_+\hat{A}_L + \om_-\hat{A}_R = \om_+UA_L + \om_-VA_R\;\;,\nl
\Bbar &=& \hat{\Bbar}_L\om_- + \hat{\Bbar}_R\om_+ = \Bbar_LU^\dag\om_- + \Bbar_RV^\dag\om_+\;\;.
\eqa
We have now combined pairs of massless chiral fermions into massive\footnote{It is of course possible that
$m_a=0$ for some $a$, especially if barren fermions have to be used.} Dirac fermions, and we have also found how to rewrite
active vertices, for instance
\bqa
\diag{171}{1.5}{-.4}&\to& i\left[\om_+(UQ^w_LU^\dag)^a_b + \om_-(VQ^w_RV^\dag)^\da_\db\right]\g^\mu\;\;,\nl
\diag{172}{1.5}{-.4}&\to& -{i\over v}\left(\om_+(UYV^\dag)^a_\db + \om_-(VY^\dag U^\dag)^\db_a\right) 
= -i{m_a\over v}\de_{ab}\;\;,
\eqa
the latter result coming from the singular-value decomposition, \eqn{singvaldec}. This proves that the Higgs is indeed
purely scalar, with no pseudoscalar component. 
The resultant form of the fermion-vector coupling depends, of course, on the model, whereas the fermion-Higgs coupling
is universal.

\section{Inclusion of fermions: example models}
\subsection{The Abelian Higgs model}
In the AH model, we restrict ourselves to $N_t=2$, in view of the discussion in (\ref{ahvertices}).
We can dispense with the superscript $w$, and we have $f^n_k = eS^n_k$. 
Since $Q$ is hermitian, we can diagonalize it, so that we use
\bq
(Q_L)^a_b = q_L^a\de_{ab}\;\;,\;\;(Q_R)^\da_\db = q_R^\da\de_{\da\db}\;\;\;\nosum\;\;.
\eq
\eqn{QKalgebra} can then be cast in the form
\bq
[Q,[Q,K^n]] = e^2K^n\;\;\;\Rightarrow\;\;\;(q_L^a-q_R^\db)^2(\ka_n)^a_\db = e^2(\ka_n)^a_\db\;\;\;\nosum\;\;.
\eq
We can simply deduce that for all values $a$ and $\db$ for which ${(\ka_n)^a}_\db$ does not vanish, 
all $q_L^a$ must be equal, or all $q_R^\db$ must be equal, or both;\footnote{If the $R$ sector, say, 
contains barren fermions we must have $q_R=0$. That is indeed the case in the Weinberg-Salam model,
where the barren fermions are neutrinos.}
we choose the latter option.
Sectors in fl-space that are not connected by nonzero $\ka_n$ entries are independent.\footnote{Think of
the lepton and quark sectors of the Standard Model. The quark sector has no barren fermions.} 
Let us concentrate on one such sector.
Here, the $Q_{L,R}$ matrices are proportional
to the unit matrix, and we find immediately that
\bq
\diag{171}{1.5}{-.4} = {i\over2}\left(\wph(q_L^a+q_R^a) + (q_L^a-q_R^a)\g^5\right)\g^\mu\,\de_{ab}\;\;.
\eq
This form of the AH model is unavoidably parity-violating, since $q_L^a-q_R^a$ cannot vanish.
Furthermore, since $(q_a-q_\db)(\ka_1)^a_\db = ie(\ka_2)^a_\db$, the matrix $Y$ is equal to
$v\ka_1$ up to a complex phase, which is taken care of by absorbing it into $U^\dag$. The fermion masses
are therefore independent of $\kt{x}$, as desired.\\

A discussion of the next-simplest model, the Apollo model, involves a considerable amount of detail special to that
model alone, and we therefore defer it to appendix \ref{apollofermions}.

\leeg{For $N_t=2$, this finishes the discussion. For $N_t=4,6,\ldots$, things are more complicated. Let us assume
that we choose the vectors $\kt{y^3},\ldots,\kt{y^{N_t}}$ as in \eqn{ychoice}. We then have fermion-scalar
vertices of the following form:
\bq
\diag{191}{1.3}{-.3} = -i\left(\om_+(U\Psi_rV^\dag)^a_b + \om_-(V\Psi_r^\dag U)^a_b\right)\;\;\;,
\;\;\;\Psi_r = (y^r)_n\ka_n\;\;.\label{ffr}
\eq
Since in this model we can write
\bq
Q_L\ka_n - \ka_nQ_R = (q_L-q_R)\ka_n = if^n_\ell\ka_\ell\;\;,
\eq
we see that
\bq
\Psi_{2s} = {1\over e}y^{2s-1}_kf^k_n\ka_n =
-i{q_L-q_R\over e}y^{2s-1}_k\ka_k = -i{q_L-q_R\over e}\Psi_{2s-1}\;\;,
\eq
the assignment of $y^{2s}$ and $y^{2s-1}$ as the real and imaginary part of a single charged scalar
is still consistent since $(q_L-q_R)/e = \pm1$. On the other hand, the vertices (\ref{ffr}) are not
automatically diagonal. A simplification arises if we choose all $\ka_n$ to be proportional to some matrix $\ka$:
\bq
\ka_n = a_n\ka\;\;,
\eq
where the $a_n$ obey\footnote{$a_n$ is not a vector in tl-space since it is complex-valued while tl-space is a real vector space.}
\bq
(q_L-q_R)a_n = if^n_\ell a_\ell\;\;\;\Rightarrow\;\;\;a_na_n = 0\;\;.
\eq
Then,
\bq
\Psi_r = Z_rY\;\;\;,\;\;\;Z_r = {y^r_na_n\over x_ka_k}\;\;,
\eq
and the vertex rule becomes
\bq
\diag{191}{1.3}{-.3} = -i{m_a\over v}\left(Z_r\om_+  + Z_r^\ast\om_-\right)\de_{ab}\;\;.
\eq}

\subsection{The Higgs-Kibble model}
\subsubsection{One  fermion doublet}
We first restrict $n_f$, the number of fermions, to 2, 
so that $Q^w_L$, $Q^w_R$, and $\ka_n$ are $2\times2$ matrices. Since now we
have $h^{wxy} = 2e\vreps^{wxy}$, we can choose
\bq
Q^w_J = -e\,(O_J)^w_z\si^z\,\la_J\;\;\;(J=L,R)\;\;,
\eq
with $O_J$ an arbitrary but fixed orthogonal matrix as before, and $\la_{L,R}$ either zero or one;
we then have $(Q^w_J)^2 = e^2\la_J^2$. By applying \eqn{QKalgebra} twice, we find
\bq
(Q^w_L)^2\ka_n + \ka_n(Q^w_R)^2 - 2(Q^w_L)\ka_n(Q^w_R) = e^2\ka_n\;\;,
\eq
in other words,
\bq
2(Q^w_L)\ka_n(Q^w_R) = e^2(\la_L^2 + \la_R^2 - 1)\ka_n\;\;;
\eq
and applying {\em this\/} twice we see that $\la_L=\la_R=1$ is not possible. We therefore take $\la_L=1$, 
$\la_R=0$, so that the right-handed fermions have {\em no\/} vector interactions, and for simplicity we take $O_L=1$.
We can again bring $Y$ into diagonal form via $UYV^\dag = D$. The fermion-vector vertex then has the form
\bq
\diag{171}{1.5}{-.4} = -ie\,\om_+\g^\mu\;(U\si^wU^\dag)^a_b\;\;.
\eq
We can further streamline the model by transforming the $W$ amplitudes:
\bq
A_w^\mu\;\to\;A_j^\mu = R^w_j\,A_w^\mu\;\;,\;\;R^w_j = \half\,\tr{U\si^wU^\dag\si^j}\;\;\;(j=1,2,3)\;\;.\label{eq:Trace identity}
\eq
Using the Fierz relations for the $2\times2$ Pauli matrices:
\bqa
&&\tr{A\,\si^w}\tr{B\,\si^w} = 2\,\tr{AB} - \tr{A}\tr{B}\;\;,   \nl
&&\tr{A\,\si^wB\,\si^w} = 2\,\tr{A}\tr{B} - \tr{AB}\;\;, 
\eqa
we can show that $R$
is an orthogonal matrix, whose application does not change the Feynman rules of the vector/scalar
sector of the model; and $R^w_j(U\si^wU^\dag) = \si^j$. We then have the Feynman rule for the fermion-vector interactions:
\bq
\diag{214}{1.5}{-.4} = -ie\,\om_+\g^\mu\;(\si^j)^a_b\;\;.
\eq
The vertex with $j=3$ is flavour-conserving, and we thus recognise the corresponding vector boson
as the neutral one. Furthermore one fermion (the `up') only emits a $W^+$, and the other one (the `down')
can only emit a $W^-$:
\bq
\diag{215}{1.5}{-.4} = -ie\sqrt{2}\,\om_+\g^\mu\;\bmat{cc}0&1\\0&0\emat^a_b\;,\;\;
\diag{216}{1.5}{-.4} = -ie\sqrt{2}\,\om_+\g^\mu\;\bmat{cc}0&0\\1&0\emat^a_b\;.
\eq

\subsubsection{More fermion doublets}
We can increase the number of fermions to $n_d$ `up' and $n_d$ `down' fermions, so that $n_f=2n_d$.
We choose $Q^w_L = -e \sigma^w \otimes\one$, explicitly:
\bq
Q_L^1 = \bmat{cc}0&$-e\one$\\$-e\one$&0\emat\;,\;
Q_L^2 = \bmat{cc}0&$ie\one$\\$-ie\one$&0\emat\;,\;
Q_L^3 = \bmat{cc}$-e\one$&0\\0&$e\one$\emat
\eq
The block form refers to the `up' and `down' sectors, and $\one$ is the $n_d\times n_d$ unit matrix.
Using the representation (\ref{gmatrices}), we then have
\bq
\ka_2=-iQ_L^1\ka_1\;,\;\ka_3 = -iQ_L^3\ka_1\;,\;\ka_4=-iQ_L^2\ka_1\;\;.
\eq
Therefore,
\bq
Y = v\Si(\vx)\ka_1\;\;,\;\;\Si(\vx) = x_1 - ix_2Q_L^1 - ix_3Q_L^3 - ix_4Q_L^2\;\;,
\eq
and $\Si$ is unitary. That means that the singular-value decomposition will automatically remove all
$x$ dependence in diagonalising $Y$ (because $\Si$ will form part of $U$): the fermion masses (and mixings) are
independent of $\kt{x}$.

There is an important restriction, however: we want the vacuum to be flavour-conserving.\footnote{Since there is no
photon, there is no notion of an electrically neutral vacuum.} Therefore the $Y$ matrix should not mix $u$ and $d$ fermions:
it must have a block-diagonal form. We must therefore choose $\ka_1$ to be (a transformation of) a block-diagonal 
matrix in fl-space:
\bq
v\ka_1 = \Si(\vz)\bmat{cc}$K_u$&0\\0&$K_d$\emat\;\;,
\eq 
with $\vz$ an arbitrary unit vector.
This kind of restriction of the form of $\ka_1$ is, in fact, also present in the canonical derivation of the Standard Model,
since also there flavour-changing vacuum terms are explicitly excluded, thus forbidding precisely one-half 
of all possible Yukawa interactions. The matrices $U$ and $V$ are now chosen as
\bq
U = C\Si(\vx)^\dag\Si(\vz)^\dag\;\;,\;\;C= \bmat{cc}$C_u$&0\\0&$C_d$\emat\;\;,\;\;V = \bmat{cc}$V_u$&0\\0&$V_d$\emat\;\;,
\eq
where
\bq
C_jK_jV_j^\dag\;\;\;(j=u,d)
\eq
is precisely the singular-value decomposition of the ($n_f\times n_f$) $K$ submatrices. Finally, we define the $R^w_j$ as
in \eqn{eq:Trace identity}, only in the `tensored' form, using $C$ rather than $U$ and $\si^w\otimes\one$. Again 
considering $W^\pm$ rather than $W^{1,2}$, we arrive at the following vertices:
\bqa
&&\diag{215}{1.5}{-.4} = -ie\sqrt{2}\,\om_+\g^\mu\;\bmat{cc}0&$C_uC_d^\dag$\\0&0\emat^a_b\;,\;\;\nl
&&\diag{216}{1.5}{-.4} = -ie\sqrt{2}\,\om_+\g^\mu\;\bmat{cc}0&0\\$C_dC_u^\dag$&0\emat^a_b\;,\;\;\nl
&&\diag{217}{1.5}{-.4} = -ie\,\om_+\g^\mu\;\bmat{cc}$\one$&0\\0&$-\one$\emat^a_b\;.\label{wcouplings}
\eqa
The matrix $C_uC_d^\dag$ is, of course, the CKM matrix in the case of quarks, or the PMNS matrix if we consider 
leptons.\footnote{Which implies, of course, the existence of right-handed neutrinos.} 

\subsection{The Electroweak model}
For simplicity, we shall use $f^w=g^w$ ({\it cf\/} \eqn{gmatrices}),
and consider only, say, the quark sector of the EW model.
Extending the HK model with an additional vector boson, we let the additional vector couple to the fermions, with
$Q^0_L$ and $Q^0_R$. Since we have demanded that $h^{0wx}=0$, $Q^0_L$ must be proportional to the unit matrix.
The only way for the commutation relations
\begin{align}
[Q^0,Q^w]=0\;\;,\;\;  Q^w_L \kappa^n - \kappa^n Q^w_R = i \left(f^w\right)^n_\ell \kappa^\ell\;\;,\;\;w=0,1,2,3
  \label{eq:EW kappa commutation relation}
\end{align}
to be consistent is to have
\begin{align}
  f^0 = e'\bmat{cccc} 0 & 0 & 1 & 0 \\ 0 & 0 & 0 & 1 \\ -1 & 0 & 0 & 0 \\ 0 & -1 & 0 & 0 \emat\,,
\end{align}
which provides a stronger constraint on $f^0$ than in the scenario where only vector bosons were included;
furthermore, the `up-down' block form of $Q^0_{L,R}$ must read
\bq
  Q^0_L = -e' \bmat{cc} $a_L\otimes\one$ & 0 \\ 0 & $a_L\otimes\one$ \emat\;\;,\;\;
  Q^0_R = -e' \bmat{cc} $a_R\otimes\one$ & 0 \\ 0 & $b_R\otimes\one$ \emat
\eq
with $a_L-a_R = -1$ and $a_L-b_R=+1$.\footnote{The lepton sector of the EW is treated the same way, only
with different assignments of $a_{L,R}$ and $b_R$.} 
As in the model containing solely vector bosons, we take the combinations $ A_\gamma^\mu = c_\theta A_0^\mu - s_\theta A_w^\mu z_w$, $A_Z^\mu = c_\theta A_w^\mu z_w + s_\theta A_0^\mu$, and $A_W^\mu = s_w A_w^\mu$ 
with $s_w=t_w,\,r_w$.\footnote{It is unnecessary to check again strict current conservation for the amplitude $A^0$, since in the
proof of section \ref{ewmodel} the identity of the active vertex does not enter.}
Since the vector $z_w$ contains the information in which $A^{1,2,3}$ are mixed, we need to perform the same mixing in the $Q^w_L$ matrices. We do so by constructing a rotation matrix $G$, whose rows are the orthonormal vectors $z_w$, $t_w$ and $r_w$, and taking the product $\tilde{R}^w_x = R^w_y G^y_x$. Performing this rotation leaves us with the Feynman rules:
\bq
\diag{218}{1.5}{-.2} = -iQ_f\g^\mu\;\;,\;\;
\diag{219}{1.5}{-.2} =  i(v_f+a_f)\g^\mu\;\;,
\eq
with
\bqa
&&Q_u = \sth e\,(a_L+a_R+1)\;\;,\;\;Q_d = \sth e\,(a_L+b_R-1)\nl
&&v_u = {e\over\cth}\left(-\cth^2+\sth^2(a_L+a_R)\right)\;\;,\;\;a_u = -{e\over\cth}\nl
&&v_d = {e\over\cth}\left(\cth^2+\sth^2(a_L+b_R)\right)\;\;,\;\;a_d = {e\over\cth}\;\;.
\eqa
Taking into account that the $f\!fW$ coupling constant $\gw$, defined by $G_F/\sqrt{2}=\gw^2/M^2$, is given by
$\gw=e\sqrt{2}$ (from \eqn{wcouplings}), it is easily checked that these vertices are, again, precisely those of
the standard electroweak model.

\section{Conclusions and acknowledgments}
We have shown how to implement the Higgs mechanism in a purely diagrammatic way, working up from the simplest
self-interacting tachyon system to the complete electroweak model. In doing so we found that special care has to taken
with tadpole-containing diagrams in order to avoid miscounting. We also proved that all theories of the type we studied
must contain a Higgs particle \cite{Higgs}, and proved the equivalence theorem cite{HV}. The symmetry structure of
theories with vector particles arises naturally from the requirement of unitarity, rather than as  preordained.\\

The authors gratefully acknowledge many useful discussions with Oscar Boher Luna and Tom de Wilt, who 
while master students have worked on several aspects of this research.

\section{Appendices}
\subsection{Antihermitian matrices\label{antihermitian}}
A hermitian matrix has an orthonormal basis of eigenvectors. The following discussion (included here since the
result is less well-known)
 describes the analogous result
for {\em anti-}hermitian matrices, that subsume antisymmetric real matrices such as the $f^w$. 
Let $M$ be an antihermitian matrix; 
 $M^\dag M$, being by construction hermitian, has an orthonormal basis of eigenvectors. Let $\kt{a_1}$ be such an
eigenvector, normalized to unity, with eigenvalue $\la$.
If $\la=0$ then
\bq
\bak{a_1}{M^\dag M}{a_1} =\; \parallel\!\! M\kt{a_1}\!\!\parallel^2 = 0\;\;,
\eq
so that $M\kt{a_1}=0$. Any other value of $\la$ must be positive, so that we can write $\la= z^2$ with $z$ real. 
For that case we define
$\kt{b_1} = (1/z)M\kt{a_1}$.
We immediately find that $\kt{b_1}$ is also normalized to unity, and orthogonal to $\kt{a_1}$; furthermore,
$M\kt{b_1}=-M^\dag\kt{b_1}=-z\kt{a_1}$.
In the complement of the span of $\kt{a_1}$ and $\kt{b_1}$, $M^\dag M$ is again hermitian, and we can repeat the process,
to find an $\kt{a_2}$ or a pair $\kt{a_2}, \kt{b_2}$, and so on.
We find that the vectors $\kt{a_j}, \kt{b_j}$ ($j=1,2,\ldots$) are an orthonormal basis, and that $M$ can be written as
\bq
M = \suml_jz_j\left(\kt{b_j}\br{a_j} - \kt{a_j}\br{b_j}\vph\right)\;\;,
\eq
where the sum runs over all nonzero eigenvalue-square-roots
 $z_j$ and their $\kt{a_j},\kt{b_j}$ pairs. If the dimension of $M$ is odd, 
there must be at least one zero eigenvalue.

\subsection{Current conservation with external vector particles \label{externalcon}}
In the derivation of \eqn{vec3rule} we have used the fact that both the $x$ and $y$ lines are axial-gauge propagators,
transverse to $n$. That assumption fails if, say, $y$ is an on-shell line, with polarization vector $\eps(q)$, for which 
$q\cdot q = q\cdot \eps(q)=0$ but $n\cdot \eps(q)$ does not necessarily vanish.\footnote{If both
$x$ and $y$ are external, the amplitude vanishes under the handlebar.} 
In that case 
we must write
\bqa
\lefteqn{\hspace{-2cm}R_n^{\mu\al}(p)\,h^{wxy}\left(\vph \left(p^\al p^\be - p^2g^{\al\be}\right) -
 \left(q^\al q^\be - q^2g^{\al\be}\right)\right)\,\eps^\be(q)}\nl
&=& i\,h^{wxy}\eps^\mu(q) - ih^{wxy}{(n\cdot\eps(q))\over(p\cdot n)}p^\mu\;\;,
\eqa
and the handlebar rule becomes
\bq
\diag{116}{1.8}{-.5} = \diag{117}{1.9}{-.5} \;-\; ih^{wxy}{(n\cdot\eps(q))\over(p\cdot n)}\,\diag{118}{1.3}{-.2}\;\;.
\eq
The first diagram on the right fits in with the proof of current conservation for $w$, while the second term
is the handlebar for the `reduced' process, where $w$ and $y$ are stripped away. We can repeat this
process until no external vector particles are left. Therefore the proof of current conservation still holds if
a finite number of external vector particles is present.

\subsection{The $g$ matrices of the HK model and the EW model\label{derivegs}}
Let $\si_j$ be the Pauli matrices, and let us denote by $S$ the $2\times2$ matrix
\bq
S = i\si_2 = \bmat{cc}0&1\\-1&0\emat\;\;.
\eq
The discussion in appendix \ref{antihermitian} shows that we can always write, in block notation,
\bq
g^1 = e\bmat{cc}$S$&0\\0&$S$\emat\;\;,\;\;
g^2 = \bmat{cc}$a_2S$&$B$\\$-B^T$&$b_2S$\emat\;\;,\;\;
g^3 = \bmat{cc}$a_3S$&$C$\\$-C^T$&$b_3S$\emat\;\;,
\eq
with $a_{2,3}$, $b_{2,3}$, $B$ and $C$ to be determined. Let us also write $h^{wxy}=ek\,\vreps^{wxy}$,
with also $k$ to be determined. The commutator identity of \eqn{falgebra}
then implies
\bq 
a_{2,3} = b_{2,3} = 0\;\;,\;\;[S,B]=kC\;\;,\;\;[S,C] = -kB\;\;.
\eq
Thus we have
\bq
[S,[S,B]] = -k^2B\;\;\;\Rightarrow\;\;\; \si_2B\si_2 = -rB\;\;,\;\;r = -1 + k^2/2\;\;.
\eq
Therefore $B$ must be a linear combination of $\si_1$ and $\si_3$, and $r=1$. If we choose $B=e\si_1$, then $C=e\si_3$, 
and we arrive at the representation of \eqn{gmatrices}, with $k=2$.\\

For the electroweak model, the matrix $f^0$ that commutes with $g^{1,2,3}$ (and consequently with
$f^{1,2,3}$) has the general form\footnote{By explicit calculation of the commutators.}
\bq
f^0 = \bmat{cc}$e_1S$&$e_2+e_3S$\\$-e_2+e_3S$&$-e_1S$\emat\;\;,\;\;
(f^0)^2 = e_1^2+e_2^2+e_3^2 \equiv e'^2\;\;.
\eq
The numbers $e_{1,2,3}$ can be chosen freely, but $f^0$ matrices with different $e_{1,2,3}$ do not 
commute with one another. There is therefore room for only {\em one\/} extra vector besides the three
self-interacting ones in the electroweak model.

\subsection{Hermiticity from cutting rules\label{cutrule}}
Since we do not use Lagrangians or actions, the hermiticity of the $Q$ and $K$ matrices must be argued diagrammatically.
To this end we may use the Cutkosky cutting rules \cite{cutko}, that embody the unitarity of a theory.
Consider a particular diagram in Quantum Electrodynamics, 
\bq
\diag{184}{2.2}{-.45} \label{loopdiag}\;\;,
\eq
where for now the labels $a,b$ and $c$ are just for telling the fermion lines apart.
The Cutkosky rule for this diagram\footnote{If the quantum numbers of $a$ and $b$ are
equal, this is also called the Optical Theorem.} reads
\bq
\diag{185}{2.2}{-.35}\;+\;\diag{186}{2.2}{-.35}\;+\;\diag{187}{2.2}{-.35}=0\;\;,\label{cut1}
\eq
where the convention is that propagators that are cut by the shaded line are on shell, while all
momentum integrations remain. On the left of the shaded line
we have the amplitude as it stands, and on the right we have the {\em complex conjugate\/} of the {\em time-reversed\/}
amplitude \cite{RK}. \eqn{cut1} can therefore also be written as
\bq
\left(\diag{184}{2.2}{-.4}\right)\;+\;
\left(\diag{189}{1.}{-.4}\right)\left(\diag{190}{1.}{-.4}\right)^\ast 
\;+\;\left(\diag{188}{2.2}{-.4}\right)^\ast= 0\;\;.    \label{cut2}
\eq
For QED, the Cutkosky rule holds for this diagram, but the cancellation is far from trivial \cite{diagrammar}.
 Let us now replace the internal photon line in the diagram (\ref{loopdiag}) by the vector particle $w$, and
 let the electron be replaced by the fermions $a,b,c$ of our model.
The terms in \eqn{cut2} then pick up, respectively, the factors
\[
(Q^w)^a_c(Q^w)^c_b\;\;,\;\;({Q^w}^\dag)^a_c({Q^w})^c_b\;\;,\;\;({Q^w}^\dag)^a_c({Q^w}^\dag)^c_b\;\;.
\]
The only reasonable way to still have the null result is to have these three factors equal. By putting $b=a$ and summing,
we therefore have
\bq
\tr{Q^wQ^w} = \tr{{Q^w}^\dag{Q^w}^\dag} = \tr{Q^w{Q^w}^\dag}\;\;\;\Rightarrow\;\;\;\tr{(Q^w-{Q^w}^\dag)^2} = 0\;\;,
\eq
which shows that $Q^w$ must be hermitian.\footnote{For an antihermitian matrix $A$,
we have $\tr{A^2} = - \sum_{a,b}|A^a_b|^2$.} Replacing the internal photon by a scalar 
in diagram (\ref{loopdiag}) does also yields a correct cutting rule;
this shows that also $K^n$ must be hermitian.

\subsection{Fermions in the Apollo model\label{apollofermions}}
Let us define, in $L,R$-block notation,
\bq
H = \tau_nK^n = \bmat{cc}0 &$Y$\\$Y^\dag$&0\emat ,\;\;R = \bmat{cc}$U$ &0\\0 & $V$\emat\;\;
\Rightarrow\;\;RHR^\dag = \bmat{cc}0&$D$\\$D$&0\emat\;\;.
\eq
Furthermore, let
\bq
Q^\g = \g_wQ^w = \bmat{cc}$Q^\g_L$&0\\0&$Q^\g_R$\emat\;\;,\;\;
\hat{Q}^\g = RQ^\g R^\dag = \bmat{cc}$\hat{Q}^\g_L$&0\\0&$\hat{Q}^\g_R$\emat
\eq
\eqn{QKalgebra} then implies
\bq
[Q^\g,H] = iv\,\g_w\,x_n\,(f^w)^n_\ell K^\ell = iv\,\g_w(e^w)_\ell K^\ell = 0\;\;.
\eq
Therefore $\hat{Q}^\g$ commutes with $RHR^\dag$, so
\bq
\hat{Q}^\g_LD = D\hat{Q}^\g_R\;,\;\; \hat{Q}^\g_RD = D\hat{Q}^\g_L\;\;.
\eq
This means that both $\hat{Q}^\g_L$ and $\hat{Q}^\g_R$ commute with $D^2$. If 
all the fermions masses are different, this means that the $\hat{Q}^\g_{L,R}$ are also diagonal in fl-space;
if several masses are equal, we can {\em make\/} the $\hat{Q}^\g_{L,R}$ diagonal by an appropriate orthogonal transformation
in fl-space. It is then easily seen that
\bq
(\hat{Q}^\g_L)^a_a = (\hat{Q}^\g_R)^a_a\;\;\mbox{if}\;m_a\ne 0\;\;\;\;\nosum\;\;:
\eq
the left- and right-handed couplings are the same for massive fermions.
This gives us the fermion-fermion-photon vertex:
\bqa
\diag{195}{1.5}{-.4} &=&
 i(\hat{Q}^\g_L)^a_a\,\g^\mu\,\de_{ab}\;\;\mbox{if $m_a>0$}\;\;\;\;\nosum\;\;,\nl
&=& i\left[\om_+(\hat{Q}^\g_L)^a_a + \om_-(\hat{Q}^\g_R)^\da_\da\right]\g^\mu\,\de_{ab}\;\;\mbox{if $m_a=0$}\;\;\;\nosum.
\eqa
The photon's interaction with massive fermions conserves parity as it should; massless fermions can
interact with parity violation without endangering the photon's current conservation.\footnote{This sidesteps the question
of the physical viability of a massless fermion coupling to massless photons: to avoid it the
$\ka_n$ must be chosen with some care.}
The two other fermion-vector couplings,
\bq
\hat{Q}^\ro = \ro_w\,RQ^wR^\dag\;\;,\;\;\hat{Q}^\tau = \tau_w\,RQ^wR^\dag\;\;,
\eq
are not automatically current-conserving under this construction, but since these couple to massive vector bosons
that is not required anyway.\\

\subsubsection{Two fermions}
It may be helpful to study a simple example. We can assume two massless fermions of both $L$ and $R$ type,
and define
\bq
(f^w)^n_k = -e\,\eps^{wnk}\;\;,\;\;Q^w_{L,R} = -{e\over2}\si^w\;\;,\;\;\ka_n = \la\,\si^n\;\;.
\eq
In this case, $\g_w = x_w$, and we have
\bq
Q^\g_{L,R} = -{e\over2}(x_w\si^w)\;\;,\;\;Y = \la\,v\,(x_n\si^n)\;\;.
\eq
We adopt polar coordinates and write  $\{\vec{\g},\vec{\rho},\vec{\tau}\}$ 
as
\bqa
\vec{\g} &=&  (\sin(\tha)\,\cos(\phi)\;,\;\sin(\tha)\,\sin(\phi)\;,\;\cos(\tha))\;\;,\nl
\vec{\ro} &=& (\cos(\tha)\,\cos(\phi)\;,\;\cos(\tha)\,\sin(\phi)\;,\;-\sin(\tha))\;\;,\nl
\vec{\tau} &=& (-\sin(\phi)\;,\;\cos(\phi)\;,\;0)\;\;,
\eqa
which provides a right-handed orthonormal base,\footnote{In the sense that
$(x.\si)(\ro.\si)(\tau.\si) = \si^1\si^2\si^3 =i$.} and further we introduce
\bq
\vec{\si}_\pm = {1\over\sqrt{2}}\left(\vec{\ro}\pm i\,\vec{\tau}\,\wph\right)\;\;.
\eq
Now, we choose the unitary matrices $U$ and $V$ as follows:
\bqa
U &=& \bmat{cc}$e^{i\phi/2}\cos(\tha/2)$&$e^{-i\phi/2}\sin(\tha/2)$\\
                      $e^{i\phi/2}\sin(\tha/2)$&$-e^{-i\phi/2}\cos(\tha/2)$\emat\;\;,\nl
V &=& \bmat{cc}$e^{i\phi/2}\cos(\tha/2)$&$e^{-i\phi/2}\sin(\tha/2)$\\
                      $-e^{i\phi/2}\sin(\tha/2)$&$e^{-i\phi/2}\cos(\tha/2)$\emat\;\;.
\eqa
This gives us
\bq
UYV^\dag = VYU^\dag = \la\,v\,\mbox{\bf{1}}\;\;\;,\;\;\;
UQ^\g_LU^\dag = VQ^\g_RV^\dag = \bmat{cc}$-e/2$&0\\0&$e/2$\emat\;\;.
\eq
We end up with two massive Dirac fermion of mass $\la v$
(independent of $\kt{x}$), one with electric charge $-e/2$ and the other with $+e/2$.
Turning to charged $W$ bosons, we find
\bqa
&&(\hat{Q}^+_L)^a_b = (U(\si_{+w}Q^w_L)U^\dag)^a_b = \bmat{cc}0&$e/\sqrt{2}$\\0&0\emat =
- (V(\si_{+w}Q^w_R)V^\dag)^\da_\db = -(\hat{Q}^+_R)^\da_\db\;\;,\nl
&&(\hat{Q}^-_L)^a_b = (U(\si_{-w}Q^w_L)U^\dag)^a_b = \bmat{cc}0&0\\$e/\sqrt{2}$&0\emat = 
-(V(\si_{-w}Q^w_R)V^\dag)^\da_\db = -(\hat{Q}^-_R)^\da_\db\;\;.\nl
\eqa
As expected, only one of the fermions can emit a $W^+$, and the other can only emit a $W^-$.
To recover the Feynman rules, we consider the emission of a $W^+$ from a fermion line:
\bq
\diag{209}{2}{-.3} = i\left(B_a\om_+\g^\mu(\hat{Q}^+_L)^a_bA^b + B_\da\om_-\g^\mu(\hat{Q}^+_R)^\da_\db A^\db\right)\;\;.
\eq
The couplings of the $W^\pm$ are seen to be purely axial in this model; $A^2$ refers to the emission of a fermion of charge
$+e/2$, and $B_1$ to the absorption of a charge $-e/2$ fermion; and of course $A^1$ emits the negative, while
$B_2$ absorbs the positive fermion. The Feynman rules are seen to be
\bq
\diag{210}{1.5}{-.3} = i{e\over\sqrt{2}}\g^5\g^\mu = \diag{211}{1.5}{-.3}\;\;.
\eq
These charge assignments then also automatically lead to the fermion-photon Feynman rules
\bq
\diag{213}{1.5}{-.3} = i{e\over2}\g^\mu\;\;\;,\;\;\;\diag{212}{1.5}{-.3} = -i{e\over2}\g^\mu\;\;.
\eq

\subsubsection{Three fermions}
As another example,\footnote{In this section, the dotted-undotted index distinction becomes really useful.} 
we choose a model with three massless fermions of both $L$ and $R$ type, and define
\bq
(f^w)^a_b = -e\,\vreps^{wab}\;\;,\;\;(Q^w_J)^a_b = ie\,\vreps^{wab}\;\;,\;\;(\ka_n)^a_\db = \la\,\vreps^{nab}\;\;,
\eq
witj $J=L,R$ as before; it is easy to verify that these satisfy \eqns{fermcurcon} with $h^{wxy}=e\,\vreps^{wxy}$.
Now $\g_w=x_w$, and we again choose the vectors $\vec{g} = \vx, \vec{\ro}$, and $\vec{\tau}$
according to 
\bq
\tau_a = \vreps^{abc}x_b\,\ro_c\;\;,
\eq
and define $\si_\pm$ as in the previous section.
We find
\bqa
&&Y^a_\db = v\,x_n(\ka_n)^a_\db = \la v(\ro^a\tau_\db - \tau^a\ro_\db),\;\;
(Q^\g_J)^a_b = x_w(Q^w_J)^a_b = ie(\ro^a\tau_b - \tau^a\ro_b)\;\;,\nl
&&(Q^\ro_J)^a_b = \ro_w(Q^w_J)^a_b = ie(\tau^ax_b - x^a\tau_b),\;\;
(Q^\tau_J)^a_b = \tau_w(Q^w_J)^a_b = ie(x^a\ro_b - \ro^ax_b).\nl
\eqa
The appropriate choices for $U$ and $V$ are
\bqa
U^a_b &=& x^ax_b + \ro^a\ro_b - \tau^a\tau_b\;\;,\nl
V^\da_\db &=& i\nu x^\da x_\db + \ro^\da\tau_\db + \tau^\da\ro_\db\;\;,\;\;\nu = \pm\;\;.\label{nuchoice}
\eqa
Note that two alternative forms for $V$ are available.
The matrix $D$ now takes the form
\bq
D^a_\db = (UYV^\dag)^a_\db = \la v(\ro^a\ro_\db + \tau^a\tau_\db) = 
\la v(\si_+^a\si_{-\db} + \si_-^a\si_{+\db})\;\;.
\eq
We have two fermions of mass $\la v$, plus one massless fermion. The fermion-photon interactions are
\bqa
&&(\hat{Q}^\g_L)^a_b = (UQ^\g_LU\dag)^a_b = e(\si_+^a\si_{-b} - \si_-^a\si_{+b})\;\;,\nl
&&(\hat{Q}^\g_R)^\da_\db = (VQ^\g_LV\dag)^\da_\db = e(\si_+^\da\si_{-\db} - \si_-^\da\si_{+\db})\;\;.
\eqa
We are led to define the off-shell amplitudes for a neutral, a positively charged, and a negatively charged
fermion as follows:
\bqa
&&\diag{196}{1}{-.1} = (x\cdot A)\;\;,\;\;\diag{197}{1}{-.1} = (\si_-\cdot A)\;\;,\;\;\diag{198}{1}{-.1} = (\si_+\cdot A)\;\;,\nl
&&\diag{199}{1}{-.1} = (B\cdot x)\;\;,\;\;\diag{200}{1}{-.1} = (B\cdot \si_+)\;\;,\;\;\diag{201}{1}{-.1} = (B\cdot \si_-)\;\;,\nl
\eqa
From
\bqa
\diag{202}{2}{-.3} &=&
 i\left(B_a\om_+\g^\mu(\hat{Q}^\g_L)^a_bA^b + B_\da\om_-\g^\mu(\hat{Q}^\g_R)^\da_\db A^\db\right)\nl
 &=& ie\left((B.\si_+)\g^\mu(\si_-.A) - (B.\si_-)\g^\mu(\si_+.A)\right)
\eqa
we derive the Feynman rules
\bq
\diag{203}{1.4}{-.2} = ie\g^\mu\;\;,\;\;\diag{204}{1.4}{-.2} = -ie\g^\mu\;\;.
\eq
The massless fermion is also the neutral one. For the fermion-$W^\pm$ vertices we have to take
\bqa
(\hat{Q}^+_L)^a_b &=& (U(\si_{+w}Q^w_L)U^\dag)^a_b = -e(x^a\si_{-b}-\si_-^ax_b)\;\;,\nl
(\hat{Q}^+_R)^\da_\db &=& (V(\si_{+w}Q^w_R)V^\dag)^\da_\db = \nu e(x^\da\si_{-\db}+\si_-^\da x_\db)\;\;,\nl
(\hat{Q}^-_L)^a_b &=& (U(\si_{-w}Q^w_L)U^\dag)^a_b = e(x^a\si_{+b}-\si_+^ax_b)\;\;,\nl
(\hat{Q}^-_R)^\da_\db &=& (V(\si_{-w}Q^w_R)V^\dag)^\da_\db = \nu e(x^\da\si_{+\db}+\si_+^\da x_\db)\;\;,
\eqa
and in the same way as above we then arrive at the following Feynman rules:\footnote{The fermion charges are 
counted along the fermion lines; the $W$ charge is counted outgoing.}
\bqa
&&\diag{205}{1.4}{-.2} = -ie\Ga_{-\nu}\g^\mu\;\;,\;\;\diag{206}{1.4}{-.2} = ie\Ga_{\nu}\g^\mu\;\;,\nl
&&\diag{207}{1.4}{-.2} = ie\Ga_{\nu}\g^\mu\;\;,\;\;\diag{208}{1.4}{-.2} = -ie\Ga_{-\nu}\g^\mu\;\;,
\eqa
where
\bq
\Ga_+ = 1\;\;\;,\;\;\;\Ga_- = \g^5\;\;.
\eq
These couplings are purely vector and axial-vector in character, but they depend on the choice of $\nu$ in
\eqn{nuchoice}: two seemingly different models that, however, are based on the same underlying physics.
Note that the two alternatives are simply related by simultaneously performing $\nu\to-\nu$
and $e\to-e$.

\end{document}